\documentclass[12pt,envcountsame]{llncs}
\usepackage{amssymb}
\usepackage{jb}

\renewcommand{\arraystretch}{1.2}

\newcommand{\3}{\ss}

\newcommand{\notion}[1]{{\em#1\/}}

\newcommand{\raa}{\longrightarrow}
\newcommand{\Ra}{\Rightarrow}
\newcommand{\La}{\Leftarrow}
\newcommand{\Lra}{\Leftrightarrow}

\renewcommand{\leq}[0]{\leqslant}
\renewcommand{\geq}[0]{\geqslant}
\newcommand{\sortdef}{::=}
\newcommand{\corresponds}{\mathrel{\mbox{\^=}}}

\newcommand{\bi}{\begin{itemize}}
\newcommand{\ei}{\end{itemize}}
\newcommand{\be}{\begin{enumerate}}
\newcommand{\ee}{\end{enumerate}}
\newcommand{\bigmid}{\rule[-0.07cm]{0.04cm}{0.4cm}\hspace*{0.1cm}}
\renewcommand{\.}[1]{\!#1\!}

\newcommand{\eqr}[1]{\ref{#1}}
\newcommand{\eqR}[1]{\ref{#1}}
\newcommand{\eqd}[1]{\label{#1}}

\newcommand{\url}[1]{{\tt #1}}

\newcommand{\NT}{{\cal N}}		
\newcommand{\G}{{\cal G}}		
\newcommand{\mR}{{\cal R}}		
\newcommand{\T}{{\cal T}}		
\renewcommand{\L}{{\cal L}}		

\newcommand{\D}{{\sf D}}		

\newcommand{\N}{I\!\!N}			

\newcommand{\pfundef}[3]{%
\newcommand{#1}[1]{#2##1#3}%
}

\pfundef{\set}{\{}{\}}			
\pfundef{\tpl}{\langle}{\rangle}	

\newcommand{\wg}[1]{\overline{#1}}	

\newcommand{\nfundef}[2]{%
\newcommand{#2}{{\sf #1}}%
}

\nfundef{wg}{\WG}			
\nfundef{hg}{\hg}			
\nfundef{sz}{\sz}			

\newcommand{\mL}{{\cal L}}		
\renewcommand{\O}{{\cal O}}

\newcommand{\abs}{\#}			

\renewcommand{\:}[4]{%
	{%
	\renewcommand{\:}[4]{%
		{%
		\renewcommand{\:}[4]{error\error}%
		\renewcommand{\j}{{##2}}%
		{##4}%
		##1...##1%
		\renewcommand{\j}{{##3}}%
		{##4}%
		}%
	}%
	\renewcommand{\i}{{#2}}%
	{#4}%
	#1\ldots#1%
	\renewcommand{\i}{{#3}}%
	{#4}%
	}%
}

\renewcommand{\,}[3]{\:{,}{#1}{#2}{#3}}

\spnewtheorem{algorithm}[theorem]{Algorithm}{\bfseries}{\rmfamily}

\newlength{\thmskip}
\newcommand{\THM}[3]{%
	\vspace{\thmskip}
        \begin{#1}%
        \Ite{#2}{ (#2) }{ }%
        $\;$\\%
        {#3}%
        \end{#1}%
	\vspace{\thmskip}
        }
\newcommand{\PRF}[2]{%
        \begin{proof}%
        {#2}%
        \end{proof}%
	\vspace{\thmskip}
        }

\newcommand{\ALGORITHM}[1] {\THM{algorithm}{}{#1}}
\newcommand{\COROLLARY}[1] {\THM{corollary} {}{#1}}
\newcommand{\LEMMA}[1]     {\THM{lemma}     {}{#1}}

\newcommand{\DEFINITION}[1]{\THM{definition}{}{#1}}
\newcommand{\PROOF}[1]     {\PRF{\em Proof. }       {#1}}
\newcommand{\pROOF}[1]     {}

\newcommand{\yY}{x}
\newcommand{\Ff}{F}
\newcommand{\Dd}{D}

\renewcommand{\l}{l}
\newcommand{\h}{h}

\renewcommand{\j}{j\hspace*{0.04cm}}
\newcommand{\q}{q}
\newcommand{\p}{p\hspace*{0.04cm}}
\newcommand{\g}{a}

\newcommand{\FF}[1]{F^{(#1)}}
\newcommand{\MM}[1]{M^{(#1)}}
\newcommand{\DD}[1]{D^{(#1)}}
\newcommand{\yy}[2]{x_{#1}^{(#2)}}
\newcommand{\xx}[2]{x_{#1}^{(#2)}}

\newcommand{\Na}{N}
\newcommand{\Nb}{N'}

\newcommand{\FFp}[1]{F'^{(#1)}}
\newcommand{\MMp}[1]{M'^{(#1)}}
\newcommand{\DDp}[1]{D'^{(#1)}}
\newcommand{\yyp}[2]{x_{#1}'^{(#2)}}

\newcommand{\?}[1]{{\LARGE\bf ??} #1}

\begin{document}




{\scriptsize\sl FIRST Technical Report 001 \hfill \today}

\begin{center}
{\Large\bf Weight Computation of Regular Tree Languages}

\vspace{1ex}

{\normalsize\rm Jochen Burghardt}

{\footnotesize\rm jochen@first.fhg.de}

\end{center}

\begin{abstract}
We present a general framework to define an application--dependent
weight measure on terms that subsumes e.g.\
total simplification orderings,
and an $\O(n \cdot \log n)$ algorithm for the simultaneous computation
of the minimal weight of a term in
the language of each nonterminal of a regular
tree grammar, based on Barzdins' liquid--flow technique.


\vspace{1ex}

{\bf Keywords.}
Regular tree language, 
Term weight, 
Minimal term,
Simplification ordering
\end{abstract}

%
%
%
%



\section{Introduction}
\label{Introduction}

Regular tree grammars or automata
\cite{Thatcher.Wright.1968}
are a generalization of regular (word) grammars allowing the description
of infinite sets of terms, aka.\ trees.
The set of regular tree languages is closed wrt.\ boolean operations
like intersection and complement, language equivalence and the
sublanguage property are decidable.
They are an important tool in various areas of computer science.

Emmelmann
\cite{Emmelmann.1991,Emmelmann.1994}
used them in a compiler generator to compute optimal target code
depending on the instruction set of the target machine, employing a
notion of weighted language membership to reflect instruction execution
costs.
Aiken and Murphy
\cite{Aiken.Murphy.1991b}
exploited the equivalence between regular tree languages and systems of
linear set constraints
in a type checking algorithm for a functional programming language.
In
\cite{Burghardt.1995b}
we used regular tree grammars to compute simple invariants of data types
that are needed in refinement verification and synthesis.
McAllester
\cite{McAllester.1992}
represented congruence classes induced by non--orientable equations in
term rewriting by regular tree languages and gave algorithms to rewrite
grammars instead of terms.
Based on an almost
similar representation,
Heinz
\cite{Heinz.1994a,Heinz.1995}
computed complete sets of term generalizations wrt.\ an equational
background theory \notion{(E--anti--unification)}.
Comon
\cite{Comon.1990}
used regular tree languages to describe sets of ground
constructor terms as
sorts, and the corresponding automaton constructions to implement sort
operations.
He provided a transformation system to
decide first--order formulas with
equality and sort membership as the only predicates.
He showed the decidability of inductive reducibility as an application.

Comon
\cite{Comon.1989} pointed out the equivalence between regular tree
automata and \notion{elementary sorted signatures}, or linear
\notion{term declarations}
by Schmidt-Schau\3 \cite{Schmidt.1988};
Uribe \cite{Uribe.1992} showed the equivalence to
linear \notion{set constraints}; Bachmair et.\ al.\
\cite{Bachmair.Ganzinger.Waldmann.1993} showed the equivalence to
the \notion{monadic class}, i.e.\ the class of first--order predicate
logic formulas with arbitrary quantifiers, with unary predicates only,
and without function symbols.

In many of the above applications, it is necessary to enumerate
terms of some computed tree language in order of increasing height, or,
more generally, some measure of \notion{weight}
which is defined dependent
on the environment the grammar algorithms are used in.
To this end, it is necessary to compute the minimal height, or weight,
of each nonterminal's language.
As a by--product, this computation also determines which nonterminals
produce the empty language, which enables to simplify the grammar
accordingly;
Aiken and Murphy
\cite{Aiken.Murphy.1991b}
found out that in their application area
this optimization had the largest impact on
practical run times at all.

In 1991, Barzdin and Barzdin \cite{Barzdin.Barzdin.1991} proposed their
\notion{liquid--flow algorithm}
which takes an incompletely given finite algebra
and acquires \notion{hypotheses} about what are probable
axioms,
using a rather involved technique including labeled graphs with several
kinds of nodes and arcs.
Recently
\cite{Burghardt.2002b},
we gave a rational reconstruction of this work that is
based on well--known algorithms on regular tree grammars.
It revealed that the core of
the liquid--flow algorithm in fact amounts to a classical fixpoint
algorithm to compute the minimal term heights of all
languages generated by nonterminals simultaneously.
However, the version of Barzdin and Barzdin has only linear
time complexity while the classical algorithm
\cite[Sect.4]{Aiken.Murphy.1991b} is quadratic.

In this paper, we generalize the liquid--flow algorithm to compute
used--defined weights instead of heights.
In the Sect.~\ref{Grammars and weights},
we present a rather general framework to define an
application--dependent weight measure on terms that subsumes
e.g.\ any total
simplification ordering \cite[Sect.~5.2]{Dershowitz.Jouannaud.1990a}
as a special case.
In Sect.~\ref{Simple Fixpoint Algorithm},
we show that a naive transfer of Barzdins' liquid--flow technique
leads to a sub--quadratic weight computation algorithm only under very
restrictive assumptions.
In Sect.~\ref{Lazy Propagation Algorithm},
we present an improved algorithm for our framework
that has a time complexity of $\O(n \cdot \log n)$ without any
restrictions.
We give correctness and complexity proofs for all presented algorithms.
%

\section{Grammars and weights}
\label{Grammars and weights}

We assume familiarity with terms and (word) grammars.

Let $\Sigma$ be
a finite set of function symbols, together with their
arities.
Let $ar$ denote the maximal arity of any function symbol in $\Sigma$.
We abbreviate the set of all $n$--ary functions by $\Sigma_n$,
and the set of all at least $n$--ary functions by $\Sigma_{\geq n}$.
Let $\T$ denote the set of ground terms over $\Sigma$,
defined in the usual way.
Let $\NT$ be a finite
set of nonterminal symbols;
we use $nt := \abs{\NT}$ to denote its cardinality.

A regular tree grammar
is a triple $\G = \tpl{\Sigma,\NT,\mR}$,
where
$\mR$ is a finite set of production rules of the form
$N \sortdef \:{\mid}1m{f_\i( \,1{n_\i}{N_{\i\j}} )} ,$
where $N, N_{ij} \in \NT$ and $f_i \in \Sigma_{n_i}$.
If some $n_i$ is zero, $f_i$ is a constant;
if $m$ is zero, $N$ produces the empty language.
We call $f_i( \,1{n_i}{N_{i\i}} )$ an alternative of the rule for
$N$.
Let $al$ denote the total number of alternatives in the grammar $\G$;
we use it as a size measure for $\G$.

For each nonterminal $N \in \NT$, exactly
one defining rule in $\mR$ must exist with $N$ as its left--hand side.
Given a fixed grammar $\G$ and a nonterminal $N \in \NT$,
the language produced by $N$,
viz.\ $\mL(N) \subseteq \T$, is defined in the usual way
as the set of all
terms derivable from $N$ as the start symbol.

Let $\D$ be a set and $(<)$ an irreflexive, total, and
well--founded\footnote{%
        I.e., each non--empty subset $S$ of $\D$ contains a minimal
        element $\min S \in S$.
	We additionally define $\min \set{} := \infty$.
	}
order on $\D$ with a maximal element $\infty$.
Let $(\leq)$ denote the reflexive closure of $(<)$.
A function $\wg{f}: \D^n \raa \D$ is called
	monotonic and increasing
	iff
	$$(\bigwedge_{i=1}^n x_i \leq y_i)
	\Ra \wg{f}( \,1n{x_\i} ) \leq \wg{f}( \,1n{y_\i} )
	\mbox{ ~ and ~ }
	\bigwedge_{i=1}^n x_i \leq \wg{f}( \,1n{x_\i} ) ,$$
respectively.
$\wg{f}$ is called a weight function iff
it is monotonic and increasing.
Let $\D$ be as above
and a signature $\Sigma$ be given.
For each $n \in \N$ and $f \in \Sigma_n$,
let a weight function $\wg{f}: \D^n \raa \D$ be given.
Define
$\WG: \T \raa \D$ inductively by
$\WG(f( \,1n{t_\i} )) := \wg{f}( \,1n{\WG(t_\i)} ) .$
For $T \subseteq \T$,
define
$$\WG(T) := \min \set{ \WG(t) \mid t \in T }. $$
Note that
$\WG(T) \in \D$ is always well--defined and
$\WG(T) = \WG(t)$ for some $t \in T$.
If $N$ is a nonterminal of a given regular tree grammar over the
signature $\Sigma$, we additionally define $\WG(N) := \WG(\mL(N))$.
A term $t \in \mL(N)$ is called minimal wrt.\ $N$ if $\WG(t) = \WG(N)$.
We assume that $\wg{f}(\,1n{x_\i})$ can always
be computed in time $\O(n)$.

The most familiar examples of weight measures are
the size $\sz(t)$, and the height $\hg(t)$ of a term $t$,
i.e.\ the total number of nodes,
and the length of the longest path from the root to any leaf,
respectively.
If $\D := \N \cup \set{ \infty }$
and $\wg{f}( \,1n{x_\i} ) := 1 + \:+1n{x_\i}$
for each $f \in \Sigma_n$,
we get $\WG(t) = \sz(t)$;
the definitions $\wg{f}( \,1n{x_\i} ) := 1 + \max \set{ \,1n{x_\i} }$
for $f \in \Sigma_n$
yield $\WG(t) = \hg(t)$.

For a more pretentious example,
let $3 \in \Sigma_0$ and $(+),(\cdot) \in \Sigma_2$,
and consider the term $t'(x,y) := x + 3 \cdot y$ with the set of proper
subterms
$S := \set{X, 3, 3 \cdot X }$.
Let $\D := (\N \times S) \cup \set{\infty}$.
The following definitions of weight functions lead to
$\WG(t)$ being a pair of the number of occurrences of $t'$ in $t$ and
the largest term in $S$ occurring at the root of $t$.
Let $i_j \in \N$ and $t_j \in \set{ X, 3, 3 \cdot X }$;
we use infix notation for weight functions as well
and assume $\wg{f}(\ldots,\infty,\ldots) := \infty$ for all $f$.
$$\begin{array}[t]{@{}rclll@{}}
& \wg{3} && := \tpl{0,3}	\\
\tpl{i_1,3} & \wg{\cdot} & \tpl{i_2,t_2}
	& := \tpl{i_1 + i_2,3 \cdot X}	\\
\tpl{i_1,t_1} & \wg{\cdot} & \tpl{i_2,t_2}
	& := \tpl{i_1 + i_2,X}
	& \mbox{if } t_1 \neq 3	\\
\tpl{i_1,t_1} & \wg{+} & \tpl{i_2,3 \cdot X}
	& := \tpl{i_1 + i_2 + 1,X} \\
\tpl{i_1,t_1} & \wg{+} & \tpl{i_2,t_2}
	& := \tpl{i_1 + i_2,X}
	& \mbox{if } t_2 \neq 3 \cdot X	\\
\multicolumn{3}{@{}c}{\wg{f}(\,1n{\tpl{i_\i,t_\i}})}
	& := \tpl{\:+1n{i_\i},X}
	& \mbox{for all other } f \in \Sigma	\\
\end{array}$$
This example can be generalized to an arbitrary linear term
$t'(\,1n{x_\i})$.

The following Lemma characterizes the weight measures $\WG$ that can be
defined by appropriate weight functions in our framework.

\LEMMA{
\eqd{25a}
Let $\D$, $(<)$, and a mapping $\phi: \T \raa \D$ be given.
There exist weight functions $\wg{f}: \D^n \raa \D$
such that $\forall t \in \T: \; \WG(t) = \phi(t)$
iff
\begin{enumerate}
\item $\phi(t') \leq \phi(t)$ if $t'$ is a subterm of $t$, and
\item $\phi(t_1) \leq \phi(t_2)
	\Ra \phi(f(\ldots,t_1,\ldots)) \leq \phi(f(\ldots,t_2,\ldots))$.
\end{enumerate}
}
\PROOF{
$\;$	\\
``$\Ra$'':
~
Let $\WG(t) = \phi(t)$.
~
Let $t'$ be a subterm of $t = c[t']$;
then $\WG(t') \leq \WG(t)$
follows from increasingness of weight functions by induction on
the height of the context $c[\cdot]$.
~
Property
2.\ follows directly from monotonicity.

``$\La$'':
\begin{itemize}
\item First note that by 2.\ and the symmetry of $(\leq)$,
        $t_1 \sim t_2 :\Lra \phi(t_1) = \phi(t_2)$
        defines a congruence relation on $\T$.
        \\
        We denote by $[t] \in \T/_\sim$
        the congruence class of $t \in \T$.
        \\
        We thus obtain an injective mapping
        $\phi^*: (\T/_\sim) \raa \D$
        defined by $\phi^*([t]) := \phi(t)$.

        In the following, we may thus assume w.l.o.g.\
        that $\T/_\sim \subseteq \D$,
	\\
        i.e.\ $\phi^*([t]) = [t] = \phi(t)$.

        Assumptions 1.\ and 2.\ then read:
        \begin{enumerate}
        \item[1'.] $[t'] \leq [t]$ if $t'$ is a subterm of $t$, and
        \item[2'.] $[t_1] \leq [t_2]
                \Ra [f(\ldots,t_1,\ldots)] \leq [f(\ldots,t_2,\ldots)]$.
        \end{enumerate}
\item For $\,1n{x_\i} \in \D$,
        define
        $\wg{f}(\,1n{x_\i}) 
        := \min \set{ [f(\,1n{t_\i})] 
        \mid \bigwedge_{i=1}^n \; x_i \leq [t_i]}$.
\item We have
        $\wg{f}(\,1n{[t_\i]}) = [f(\,1n{t_\i})]$
        for all $\,1n{t_\i} \in \T$:

        $\wg{f}(\,1n{[t_\i]}) \leq [f(\,1n{t_\i})]$
        is obvious, since $\bigwedge_{i=1}^n \; [t_i] \leq [t_i]$.

        For any $t'_i$ with $[t_i] \leq [t'_i]$
        we have by repeated application of
        2'.\ that
        $[f(\,1n{t_\i})] \leq [f(\,1n{t'_\i})]$.

        Hence,
        $[f(\,1n{t_\i})]
        \leq \min \set{ [f(\,1n{t'_\i})]
        \mid \bigwedge_{i=1}^n \; [t_i] \leq [t'_i]}
        = \wg{f}(\,1n{[t_\i]})$.
\item Each $\wg{f}$ is increasing:
        \\
        \begin{tabular}[t]{@{}ll@{\hspace*{1cm}}l@{}}
        & $\wg{f}(\,1n{x_\i})$  \\
        $=$ & $\min \set{ [f(\,1n{t_\i})]
                \mid \bigwedge_{i=1}^n \; x_i \leq [t_i]}$
                & Def.~$\wg{f}$ \\
        $\geq$ & $\min \set{ t_j
                \mid \bigwedge_{i=1}^n \; x_i \leq [t_i]}$
                & by 1'.	\\
        $\geq$ & $x_j$  \\
        \end{tabular}
\item Each $\wg{f}$ is monotonic:
        \\
        Let $\bigwedge_{i=1}^n \; x_i \leq y_i$,
        then:
        \\
        \begin{tabular}[t]{@{}ll@{\hspace*{1cm}}l@{}}
        & $\wg{f}(\,1n{x_\i})$  \\
        $=$ & $\min \set{ [f(\,1n{t_\i})]
                \mid \bigwedge_{i=1}^n \; x_i \leq [t_i]}$
                & Def.~$\wg{f}$ \\
        $\leq$ & $\min \set{ [f(\,1n{t_\i})]
                \mid \bigwedge_{i=1}^n \; y_i \leq [t_i]}$
                & since 
		$\bigwedge_{i=1}^n \; y_i \leq [t_i]
                \Ra \bigwedge_{i=1}^n \; x_i \leq [t_i]$   \\
        $=$ & $\wg{f}(\,1n{y_\i})$ & Def.~$\wg{f}$      \\
        \end{tabular}
\item $\WG(t) = [t]$ for all terms $t$
        \\
        follows immediately by induction on $t$.
\end{itemize}
\qed
}
\pROOF{
``$\Ra$'':
Straight forward, using $\phi(t) = \WG(t)$.
\\
``$\La$'':
$t_1 \sim t_2 :\Lra \phi(t_1) = \phi(t_2)$
defines a congruence relation on $\T$,
and $\T/_\sim$ is isomorphic to a subset of $\D$.
W.l.o.g.\ assume $\D \subseteq \T/_\sim$,
and define
$$\wg{f}(\,1n{x_\i})
:= \min \set{ [f(\,1n{t_\i})]
\mid \bigwedge_{i=1}^n \; x_i \leq [t_i]} ,$$
where $[t] \in \T/_\sim$ denotes the congruence class of $t \in \T$.
Show
$\wg{f}(\,1n{[t_\i]}) = [f(\,1n{t_\i})]$
for all $\,1n{t_\i} \in \T$ and monotonicity and increasingness of
$\wg{f}$ directly, and $\WG(t) = [t] = \phi(t)$ by induction on $t$.
}

As an application of Lem.~\eqr{25a},
let $\D = \T \cup \set{\infty}$,
ordered by a (reflexive) total
simplification ordering $(\leq)$
\cite[Sect.~5.2]{Dershowitz.Jouannaud.1990a},
and let $\phi$ be the identity function.
Property 1.\ and 2.\ is satisfied,
since $(\leq)$ contains the subterm
ordering and is closed under context application, respectively.
Then $\WG(t) = t$ for each term $t$,
i.e., no two distinct terms have the same weight,
and $\WG(N)$ yields the least term of $\mL(N)$ wrt.\ $(\leq)$.

In the rest of this paper, we assume a fixed given grammar
$\G = \tpl{\Sigma,\NT,\mR}$.
For a nonterminal $N \in \NT$,
we tacitly assume
its defining rule to be
$$N \sortdef \:{\mid}1m{f_\i( \,1{n_\i}{N_{\i\j}} )} .$$

\section{Simple Fixpoint Algorithm}
\label{Simple Fixpoint Algorithm}

First, we adapt the classical algorithm from
\cite{Aiken.Murphy.1991b} to our framework.

\ALGORITHM{ \eqd{38}
\notion{(Naive fixpoint language weight computation)}
~
Let $\xx{N}{i} \in \D$ be defined by
\begin{itemize}
\item $\xx{N}{0} := \infty$
	for each $N$,
	and
\item $\xx{N}{k+1}
	:= \min \set{ \wg{f}_i( \,1{n_i}{\xx{N_{i\i}}{k}} )
	\mid 1 \leq i \leq m }$
	for each $N$
	and $k=\,1{nt}\i$.
\end{itemize}
In the $k$'th computation cycle, all $\xx{N}{k}$ are computed from
all $\xx{N}{k-1}$ by the above formula.
When the algorithm stops after the $nt$'th cycle, we have
$\xx{N}{nt} = \WG(N)$ by Cor.~\eqR{43} below.
Since in each cycle all alternatives are evaluated, we get a time
complexity of $\O(nt \cdot al \cdot ar)$.
\qed
}

\LEMMA{ \eqd{40}
Given the settings of Alg.~\eqr{38},
we have for all nonterminals $N$ of $\G$,
all $t \in \mL(N)$,
and all $k \in \N$,
that $\hg(t) \leq k \Ra x_N^{(k)} \leq \WG(t)$.
}
\PROOF{
Induction on $k$:
\bi
\item $k = 0$: trivial, since no ground terms of height $0$ exist.
\item $k \leadsto k+1$:
        if $t = f_i( \,1{n_i}{t_\i} ) \in \mL(N)$
        has a height $\leq k+1$,
        and assuming the rule
        $N \sortdef \bigmid_{i=1}^m \; f_i( \,1{n_i}{N_{i\i}} )$,
        we have $\hg(t_j) \leq k$ for $j= \,1n\i$,
        and
        \\
        \begin{tabular}[t]{@{}l@{$\;$}l@{\hspace*{0.5cm}}l@{}}
        & $x_N^{(k+1)}$ \\
        $\leq$ & $\wg{f}_i( \,1{n_i}{x_{N_{i\i}}^{(k)}} )$
                 & Alg.~\eqr{38}        \\
        $\leq$ & $\wg{f}_i( \,1{n_i}{\WG(t_\i)} )$
                & $t_j \in \mL(N_{ij})$ for $j= \,1{n_i}\i$,
                I.H.,
                $\wg{f}_i$ monotonic    \\
        $=$ & $\WG(t)$ & Def.~$\WG$     \\
        \end{tabular}
\ei
\qed
}

\LEMMA{ \eqd{41}
Given the settings of Alg.~\eqr{38},
we have for all nonterminals $N$ that
\\
$x_N^{(k)} < \infty
\Ra \exists t \in \mL(N): \;\; \hg(t) \leq k \wedge x_N^{(k)} = \WG(t)$.
}
\PROOF{
Induction on $k$:
\bi
\item $k = 0$: trivial, since $x_N^{(0)} = \infty$.
\item $k \leadsto k+1$:
        Assume the rule
        $N \sortdef \bigmid_{i=1}^m \; f_i( \,1{n_i}{N_{i\i}} )$.
        ~
        First, we have
        \\
        \begin{tabular}[t]{@{}l@{$\;$}l@{\hspace*{0.5cm}}l@{}}
        & $\infty$      \\
        $>$ & $x_N^{(k+1)}$ & assumption        \\
        $=$ & $\min \set{ \wg{f}_i( \,1{n_i}{x_{N_{i\i}}^{(k)}} )
                \mid 1 \leq i \leq m }$
                & Alg.~\eqr{38} \\
        $=$ & $\wg{f}_l( \,1{n_l}{x_{N_{l\i}}^{(k)}} )$
                for some $l \in \set{ \,1m\i }$ & Def.~$\min$   \\
        $\geq$ & $x_{N_{lj}}^{(k)}$ for $j= \,1{n_l}\i$
                & $\wg{f}_l$ increasing \\
        \end{tabular}
        \\
        If $n_l = 0$, $f_l \in \mL(N)$ and
        $\WG(f_l)
        = \wg{f}_l
        = x_N^{(k+1)}$
        as above, and we are done.
        \\
        If $n_l > 0$,
        by induction hypothesis, for each $j=\,1{n_l}\i$
        exists some $t_j \in \mL(N_{lj})$
        with $\hg(t_j) \leq k$
        and $x_{N_{lj}}^{(k)} = \WG(t_j)$,
        \\
        hence
        $f_l( \,1{n_l}{t_\i} ) \in \mL(N)
        \wedge \hg(f_l(\,{1}{n_l}{t_\i})) \leq k+1$,
        and
        \\
        \begin{tabular}[t]{@{}l@{$\;$}l@{\hspace*{0.5cm}}l@{}}
        & $\WG(f_l( \,1{n_l}{t_\i} ))$  \\
        $=$ & $\wg{f}_l( \,1{n_l}{\WG(t_\i)} )$ & Def.~$\WG$ \\
        $=$ & $\wg{f}_l( \,1{n_l}{x_{N_{l\i}}^{(k)}} )$
                & property of the $t_j$ \\
        $=$ & $\min \set{ \wg{f}_i( \,1{n_i}{x_{N_{i\i}}^{(k)}} )
                \mid 1 \leq i \leq m }$
                & Def.~$l$      \\
        $=$ & $x_N^{(k+1)}$ & Alg.~\eqr{38}     \\
        \end{tabular}
        \\
        (Note that this induction step holds even for $k=1$,
        since $x_N^{(1)} < \infty$ requires some nullary $f_l$ in the
        rule of $N$.)
\qed
\ei
}

\LEMMA{ \eqd{42}
$\xx{N}{k} =
\min \set{ \WG(t) \mid t \in \mL(N), \; \hg(t) \leq k }$.
}
\pROOF{
By induction on $k$,
show successively:
\begin{itemize}
\item
	$\forall N \in \NT, \; t \in \mL(N), \; k \in \N: \;\;
	\hg(t) \leq k \Ra \xx{N}{k} \leq \WG(t)$,
	and
\item
	$\forall N \in \NT, \; k \in \N: \;\;
	(\xx{N}{k} < \infty
	\Ra \exists t \in \mL(N): \;\;
	\hg(t) \leq k \wedge \xx{N}{k} = \WG(t))$.
	\qed
\end{itemize}
}
\PROOF{
$\;$	\\
For $x_N^{(k)} = \infty$,
we have $\min \set{ \WG(t) \mid t \in \mL(N), \hg(t) \leq k } = \infty$
by Lemma \eqr{40}.
\\
For $x_N^{(k)} < \infty$,
let $t_0$ be the term obtained by Lemma \eqr{41}.
Then:
\\
\begin{tabular}[b]{@{}l@{$\;$}l@{\hspace*{0.5cm}}l@{}}
& $x_N^{(k)}$   \\ 
$\leq$ & $\min \set{ \WG(t) \mid t \in \mL(N), \hg(t) \leq k }$
        & Lemma \eqr{40}        \\
$\leq$ & $\WG(t_0)$
        & since $t_0 \in \mL(N), \hg(t_0) \leq k$       \\
$=$ & $x_N^{(k)}$ & Lemma \eqr{41}      \\
\end{tabular}
\qed
}

\COROLLARY{ \eqd{43}
\notion{(Correctness of Alg.~\eqr{38})}
~
$\xx{N}{nt} = \WG(N)$ for all $N$.
}
\pROOF{
Use Lem.~\eqr{42},
and apply the pumping lemma
\cite[Sect.~1.2]{Comon.Dauchet.Gilleron.1999}
to show
that a term of height greater than $nt$ cannot have minimal weight.
}
\PROOF{
$\;$	\\
If $\mL(N) = \set{}$,
we have by Cor.~\eqr{42}
that $x_N^{(k)} = \infty = \WG(N)$
for all $k$.
\\
Else, by the pumping lemma for regular tree languages
\cite[Sect.~1.2]{Comon.Dauchet.Gilleron.1999},
we can find for each term $t \in \mL(N)$ with $\hg(t) > k \geq nt$
a smaller one $t' \in \mL(N)$ with $\hg(t') \leq nt$.

Since all weight functions $\wg{f}$ are increasing,
the well--known construction of $t'$ from $t$ ensures
that $\WG(t') \leq \WG(t)$;
i.e., $t$ does not contribute to the minimum $\WG(N)$.
Hence:
\\
\begin{tabular}[b]{@{}l@{$\;$}l@{\hspace*{0.5cm}}l@{}}
& $\WG(N)$      \\
$=$ & $\min \set{ \WG(t) \mid t \in \mL(N) }$ & Def.~$\WG$      \\
$=$ & $\min \set{ \WG(t) \mid t \in \mL(N), \hg(t) \leq k }$
        & see above     \\
$=$ & $x_N^{(k)}$ & Cor.~\eqr{42}       \\
\end{tabular}
\qed
}

Next, we apply Barzdins' liquid--flow technique in a straight forward
way to Alg.~\eqr{38}.
The basic idea is to recompute an alternative only if some of its
argument values has changed, i.e.\ belongs to the \notion{water front}.

\ALGORITHM{ \eqd{49c} 
\notion{(Barzdins' liquid--flow technique)}
~
The weights of all nonterminals can be
computed in time $\O(nt \cdot al \cdot ar)$,
using the following fixpoint algorithm.
Maintain a set $F \subseteq \NT$, called \notion{water front}
in \cite{Barzdin.Barzdin.1991},
such that
$\FF{k+1} = \set{ N \in \NT \mid \yy{N}{k+1} < \yy{N}{k}}$
for all $k \in \N$.
Define
\begin{itemize}
\item $\yy{N}{0} := \infty$
        for each $N$,
\item $\yy{N}{1}
	:= \min \set{ \wg{f}_i
	\mid 1 \leq i \leq m, \; f_i \in \Sigma_0}$
	for each $N$,
        and
\item $\yy{N}{k+1}
        := \min (\set{ \yy{N}{k}} \cup
	\set{ \wg{f}_i( \,1{n_i}{\yy{N_{i\i}}{k}} )
        \mid 1 \leq i \leq m, \;
	\bigvee_{j=1}^{n_i} \; N_{ij} \in \FF{k} })$
	\\
        for each $N$
        and each $k \in \N$.
\item Stop the computation after cycle $(k+1)$, if $\FF{k+1}$ is empty.
\end{itemize}
In cycle $(1)$, the entire grammar has to be inspected once, to compute
$\yy{N}{1}$ and at the same time $\FF{1}$.
In each later cycle, we tacitly assume
that we need to inspect only those alternatives
$f_i( \,1{n_i}{N_{i\i}} )$
for which $N_{ij} \in \FF{k}$ for some $j \in \set{\,1{n_i}\i}$.

As in \cite{Barzdin.Barzdin.1991}, an appropriate pointer structure,
linking each nonterminal $N_{ij}$ (\/\notion{domain node} in
\cite{Barzdin.Barzdin.1991}) to all
alternatives \notion{(functional nodes)}
it occurs in, is assumed to have been built before cycle
$(0)$, by inspecting the whole grammar once.

Below we will show linear complexity under certain
restrictive requirements to the weight functions.
\qed
}

It is easy to see by induction on $k$ that
each $\xx{N}{k}$ has the same value in Alg.~\eqr{49c} as in
Alg.~\eqr{38}.
Since
$\FF{k} = \set{}
\Ra \yy{N}{k+1} = \yy{N}{k} \wedge \FF{k+1} = \set{}$,
the algorithm may stop in this case.

\begin{figure}
\begin{center}
\begin{tabular}[t]{@{}|@{$\;$}l@{$\;$}l@{$\;$}|@{}}
\hline
$Q_0$ & $\sortdef \g$  \\
$Q_{n+1}$ & $\sortdef q(P_{n+1}) \mid j(Q_n)$  \\
$P_{n+1}$ & $\sortdef p(Q_n)$   \\
\hline
\end{tabular}
\hfill
\begin{tabular}[t]{@{}|@{$\;$}l@{$\;$}l@{$\;$}|@{}}
\hline
$\wg{\q}(x)$ & $:= x$  \\
$\wg{\p}(x)$ & $:= 2 \cdot x$  \\
$\wg{\j}(x)$ & $:= 2 \cdot x + 1$	\\
$\wg{\g}$ & $:= 0$      \\
\hline
\end{tabular}
\hfill
\begin{tabular}[t]
	{@{}|@{$\;$}l@{$\;$}|@{$\;$}l@{$\;$}|@{$\;$}l@{$\;$}|@{}}
\hline
$\mL(Q_n)$ & $\hg$ & $\WG$	\\
\hline
$\q\p\ldots\q\p\g$ & $2 \cdot n + 1$ & $0$	\\
$\j\ldots\j\g$ & $n + 1$ & $2^n - 1$	\\
\hline
\end{tabular}
\end{center}
\caption{Example grammar, weight functions, and
	minimal / maximal terms}
\label{Grammar and weight functions in Exm.}
\end{figure}

\begin{figure}
\begin{center}
\begin{tabular}[t]{@{}|r||*{7}{r@{$\;\;$}l|}@{}}
\hline
$k$
	& \multicolumn{2}{c|}{$Q_0$}
	& \multicolumn{2}{c|}{$P_1$}
	& \multicolumn{2}{c|}{$Q_1$}
	& \multicolumn{2}{c|}{$P_2$}
	& \multicolumn{2}{c|}{$Q_2$}
	& \multicolumn{2}{c|}{$P_3$}
	& \multicolumn{2}{c|}{$Q_3$}	\\
\hline
\hline
$0$
	& \multicolumn{2}{r|}{$\infty$}
	& \multicolumn{2}{r|}{$\infty$}
	& \multicolumn{2}{r|}{$\infty$}
	& \multicolumn{2}{r|}{$\infty$}
	& \multicolumn{2}{r|}{$\infty$}
	& \multicolumn{2}{r|}{$\infty$}
	& \multicolumn{2}{r|}{$\infty$}	\\
\hline
$1$
	& $\g$ & $0$
	&&
	&&
	&&
	&&
	&&
	&&	\\
\hline
$2$
	&&
	& $\p\g$ & $0$
	& $\j\g$ & $1$
	&&
	&&
	&&
	&&	\\
\hline
$3$
	&&
	&&
	& $\q\p\g$ & $0$
	& $\p\j\g$ & $2$
	& $\j\j\g$ & $3$
	&&
	&&	\\
\hline
$4$
	&&
	&&
	&&
	& $\p\q\p\g$ & $0$
	& $\j\q\p\g$ & $1$
	&&
	&&	\\
$\;$
	&&
	&&
	&&
	&&
	& $(\q\p\j\g$ & $2)$
	& $\p\j\j\g$ & $6$
	& $\j\j\j\g$ & $7$	\\
\hline
$5$
	&&
	&&
	&&
	&&
	& $\q\p\q\p\g$ & $0$
	& $\p\j\q\p\g$ & $2$
	& $\j\j\q\p\g$ & $3$	\\
$\;$
	&&
	&&
	&&
	&&
	&&
	&&
	& $(\q\p\j\j\g$ & $6)$	\\
\hline
$6$
	&&
	&&
	&&
	&&
	&&
	& $\p\q\p\q\p\g$ & $0$
	& $\j\q\p\q\p\g$ & $1$	\\
$\;$
	&&
	&&
	&&
	&&
	&&
	&&
	& $(\q\p\j\q\p\g$ & $2)$	\\
\hline
$7$
	&&
	&&
	&&
	&&
	&&
	&&
	& $\q\p\q\p\q\p\g$ & $0$	\\
\hline
\end{tabular}
\caption{Algorithm~\eqr{38} / \eqr{49c} running on the example grammar}
\label{Algorithm 38 in Exm.}
\end{center}
\end{figure}

As an example,
let $\Sigma_0 := \set{ \g }$ and $\Sigma_1 := \set{ \q, \p, \j }$;
consider the grammar (left) and weight functions (middle)
shown in Fig.~\ref{Grammar and weight functions in Exm.}, where
$\D := \N \cup \set{ \infty }$ and $n = \,{0}{n_{\max}}\i$.
The grammar has $2 \cdot n_{\max} + 1$ rules and $3 \cdot n_{\max} + 1$
alternatives.
Since we have at most unary functions,
we can omit parentheses around function arguments in terms to enhance
readability and to indicate the connection to word grammars.

Observe that each term $t$
in $\mL(Q_n)$ can be read as a binary number in
reversed notation%
	\footnote{%
	Think of ``$\q$'' and ``$\p$'' as forming the left and right
	half of the digit ``$0$'', respectively,
	think of ``$\j$'' as of the
	digit ``$1$'', and ignore ``$\g$''.
	}%
,
e.g.\
$\mL(Q_3) \ni \q\p\j\j\g \corresponds 110$,
and that by construction
the weight $\WG(t)$ corresponds to this binary number,
e.g.\ $\WG(\q\p\j\j\g) = 6$.

In this correspondence, $\mL(Q_n)$ is the set of all binary numbers with
exactly $n$ digits;
in particular, each $\mL(Q_n)$ is finite.
Hence, both a term of maximal height and of maximal weight exists,
by construction, it has minimal weight and minimal height, respectively;
cf.~Fig.~\ref{Grammar and weight functions in Exm.} (right).

Figure~\ref{Algorithm 38 in Exm.} shows the cycles
of Alg.~\eqr{38} in computing the weights of each $Q_n$ and $P_n$,
where $n_{\max} = 3$.
For each nonterminal $N$, the right entry in its column shows the
value of $\xx{N}{k}$ while the left entry shows a term of this weight in
$\mL(N)$.
Empty entries mean that $\xx{N}{k} = \xx{N}{k-1}$.
Entries in parentheses are computed by the algorithm, but don't lead to
an update, since they are larger than their predecessors.

The algorithm stops after cycle $(7)$, since there are $7$ distinct
nonterminals in the grammar.
Since Alg.~\eqr{49c}
computes the same values as Alg.~\eqr{38},
Fig.~\ref{Algorithm 38 in Exm.} illustrates both algorithms
simultaneously.
The water front $\FF{k}$ consists of
all nonterminals having a nonempty entry in line $k$, except for $k=0$.

Next, we give a sufficient criterion for Alg.~\eqr{49c} to run in linear
time.
Define
$$\begin{array}[t]{@{}llr@{}}
\l_0 & := \min \set{ \wg{f} \mid f \in \Sigma_0},	\\
\h_0 & := \max \set{ \wg{f} \mid f \in \Sigma_0},	\\
\l(x) & := \min \set{ \wg{f}(\l_0,...,x,...,\l_0)
	\mid f \in \Sigma_{\geq 1}}, & \mbox{and}	\\
\h(x) & := \max \set{ \wg{f}(\,1nx)
	\mid f \in \Sigma_{\geq 1}}.	\\
\end{array}$$
In the definition of $\l(x)$, the minimum ranges over all argument
positions of $x$,
e.g.\ $\l(x) = \min \set{\wg{g}(x,\l_0), \wg{g}(\l_0,x)}$
if $\Sigma_{\geq 1} = \Sigma_2 = \set{g}$.
Note that the maxima in the definitions of $\h_0$ and $\h(x)$ are
well--defined since they range over a finite set each.

\LEMMA{ \eqd{43b}
In the settings of Alg.~\eqr{38},
we have for each $N \in \NT$ and each $k \in \N$
\\
that
$x_N^{(k+1)} < x_N^{(k)}$ iff 
some $t \in \mL(N)$ exists with $\hg(t) = k+1$
and $\WG(t) < \WG(t')$ for each $t' \in \mL(N)$ with $\hg(t') \leq k$.
}
\PROOF{
Follows immediately from Cor.~\eqr{42},
\\
since $\min A < \min B$
iff $\exists a \in A \.\setminus B \; \forall b \in B: \; a < b$
for arbitrary ordered sets.
%
%
%
\qed
}

\LEMMA{ \eqd{49hi}
If some $p_0 \in \N$ exists such that
$\abs{\set{d \in \D \mid \l^i(\l_0) \leq d \leq \h^i(\h_0)}}
\leq p_0$
for all $i \in \N$,
then each $\xx{N}{\cdot}$
can change its value at most $p_0$ times during the
execution of Alg.~\eqr{38}.
(Here, $\l^i(\l_0)$ denotes the $i$--fold
application of $\l$ to $\l_0$, etc.)
}
\pROOF{
Show successively:
\begin{itemize}
\item $\l(\cdot)$ and $\h(\cdot)$ are monotonic and increasing.
\item $\l(\max \set{ \,1n{x_\i} }) \leq \wg{f}( \,1n{x_\i} )
	\leq \h(\max \set{ \,1n{x_\i} })$
	for all $\,1n{x_\i}$ in the range of $\WG$
	and all $f \in \Sigma_{\geq 1}$.
\item $\l^{\hg(t)-1}(\l_0) \leq \WG(t) \leq \h^{\hg(t)-1}(\h_0)$
	for all $t \in \T$, by induction on $t$.
\item If in Alg.~\eqr{38} the value of $\xx{N}{\cdot}$
	changes at $k=\,1p{k_\i}$,
	then
	$$\l^{k_p-1}(\l_0)
	\leq \WG(t_p)
	= \:{<}p1{\xx{N}{k_\i}}
	= \WG(t_1)
	\leq \h^{k_1-1}(\h_0)
	\leq \h^{k_p-1}(\h_0)$$
	for appropriate $t_i$ obtained from Lem.~\eqr{42};
	hence $p \leq p_0$.
	\qed
\end{itemize}
}
\PROOF{
$\;$
\begin{enumerate}
\item $\l(\cdot)$ and $\h(\cdot)$ are monotonic and increasing:
        \\
        If $x \leq y$,
        then 
        $\wg{f}(\l_0,...,x,...,\l_0) \leq
        \wg{f}(\l_0,...,y,...,\l_0)$
        for all $f \in \Sigma_{\geq 1}$,
        and hence
        $\l(x) \leq \l(y)$.
        ~
        Since $x \leq \wg{f}(\l_0,...,x,...,\l_0)$
        for all $f \in \Sigma_{\geq 1}$,
        we have $x \leq \l(x)$.
        ~
        Similar for the monotonicity and increasingness of $\h$.
\item $\l_0 \leq \wg{f} \leq \h_0$ for all $f \in \Sigma_0$:
        \\
        Obvious.
\item $\l(\max \set{ \,1n{x_\i} }) \leq \wg{f}( \,1n{x_\i} )
        \leq \h(\max \set{ \,1n{x_\i} })$
	\\
        for all $\,1n{x_\i}$ in the range of $\WG$
        and all $f \in \Sigma_{\geq 1}$:
        \\
        Since each $x_i$ is in the range of $\WG$,
        we have $\l_0 \leq x_i$.
        \\
        W.l.o.g.\ let $x_1 = \max \set{ \,1n{x_\i} }$.
        ~
        Then,
        \\
        \begin{tabular}[t]{@{}ll@{\hspace*{1cm}}l@{}}
        & $\l(x_1)$     \\
        $\leq$ & $\wg{f}(x_1,\l_0,...,\l_0)$ & Def.~$\l$        \\
        $\leq$ & $\wg{f}( \,1n{x_\i} )$ & $\l_0 \leq x_i$       \\
        $\leq$ & $\wg{f}( \,1n{x_1} )$ & $x_i \leq x_1$ \\
        $\leq$ & $\h(x_1)$ & Def.~$\h$  \\
        \end{tabular}
\item $\l^{\hg(t)-1}(\l_0) \leq \WG(t) \leq \h^{\hg(t)-1}(\h_0)$
        for all $t \in \T$:
        \\
        Induction on $t$:
        \\
        If $t = f$ is a constant,
        we have
        $\l^{\hg(t)-1}(\l_0) 
        = \l_0 
        \leq \wg{f} 
        = \WG(t) 
        \leq \h_0 
        = \h^{\hg(t)-1}(\h_0)$.
        \\
        If $t = f( \,1n{t_\i} )$,
        we have $\hg(t_i) = \hg(t)-1$ for some $i \in \set{ \,1n\i }$,
	\\
        and $\hg(t_j) \leq \hg(t)-1$ for all $j=\,1n\i$.
        ~
        Hence,
        \\
        \begin{tabular}[t]{@{}ll@{\hspace*{1cm}}l@{}}
        & $\l^{\hg(t)-1}(\l_0)$ \\
        $\leq$ & $\l(\WG(t_i))$
                & since
                $\l^{\hg(t)\.-2}(\l_0)
                \.= \l^{\hg(t_i)\.-1}(\l_0) 
                \.\leq \WG(t_i)$
                by I.H. \\
        $\leq$ & $\l(\max \set{ \,1n{\WG(t_\i)} })$
                & Def.~$\max$, ~ $i \in \set{ \,1n\i }$ \\
        $\leq$ & $\wg{f}( \,1n{\WG(t_\i)} )$ & by 3.    \\
        $=$ & $\WG(t)$ & Def.~$\WG(t)$  \\
        $\leq$ & $\h(\max \set{ \,1n{\WG(t_\i)} })$ & by 3.     \\
        $\leq$ & $\h^{\hg(t)-1}(\h_0)$
                & since
                $\WG(t_j) \leq \h^{\hg(t)-2}(\h_0)$ 
		for all $j$
                by I.H. \\
        \end{tabular}
\item If in Alg.~\eqr{38} the value of $\xx{N}{\cdot}$
        changes at $k=\,1p{k_\i}$,
        then $p \leq p_0$:
        \\
        Let
        $\set{ \,1p{k_\i} }
        := \set{ k+1 \in \N \mid x_N^{(k+1)} < x_N^{(k)} }$
        and $\:{<}1p{k_\i}$.
        \\
        From Cor.~\eqr{43b}, we obtain $\,1p{t_\i} \in \mL(N)$
	\\
        such that $\hg(t_i) = k_i$
        and $\WG(t_i) = x_N^{(k_i)}$ for $i=\,1p\i$.
        \\
        From 4., we thus have
        \\
        $\l^{k_p-1}(\l_0)
        \leq \WG(t_p) 
        = \:{<}p1{x_N^{(k_\i)}}
        = \WG(t_1)
        \leq \:{\leq}1p{\h^{k_\i-1}(\h_0)}$.
        \\
        Hence, by assumption
        $p \leq p_0$.
        \qed
\end{enumerate}
}

Let us
compute the time requirements of Alg.~\eqr{49c} under the assumption
of Lem.~\eqr{49hi}.
Let $C_N$ denote the number of cycles where $\yy{N}{\cdot}$
changes its value
during the computation, i.e.\ where $\yy{N}{k+1} < \yy{N}{k}$.
From Lem.~\eqr{49hi},
we get $C_N \leq p_0$ for all $N \in \NT$
and some $p_0 \in \N$ depending not on the grammar, but only on the
choice of the weight functions.
A certain alternative
$\wg{f}_i( \,1{n_i}{\yy{{N_{i\i}}}{k}})$
is recomputed
$$\sum_{j=1}^{n_i} C_{N_{ij}}
\leq \sum_{j=1}^{n_i} p_0
\leq ar \cdot p_0$$
times during algorithm execution.
Altogether, there are
at most $al \cdot ar \cdot p_0$ re--computations of alternatives,
requiring time $\O(al \cdot ar^2 \cdot p_0)$, including updating the
respective $\yy{N}{\cdot}$.
Initialization of the pointer structure and cycles $(0)$ and $(1)$
requires time $\O(al \cdot ar + nt + al)$.
Hence, we get an overall time requirement of
$\O(al \cdot ar^2 \cdot p_0)$, i.e.\ the algorithm runs linear in the
number of alternatives.

In the binary numbers example from above,
the computation still takes $\O(n_{\max}^2)$ cycles since the
requirements of Lem.~\eqr{49hi} are not satisfied.
In general, these requirements are violated as soon as there are two
weight functions $\wg{f}$ and $\wg{g}$ such that $\wg{f}(x) < \wg{g}(x)$
for all $x \in \D$ and $\wg{g}$ is strictly monotonic%
	\footnote{%
	I.e.,
	$x < y \Ra \wg{g}(x) < \wg{g}(y)$.
	We use unary functions for simplicity;
	the argument is similar for higher arity functions.
	}%
,
since then
$$\l^i(\l_0)
\leq \wg{f}^i(\l_0)
< \wg{g}(\wg{f}^{i-1}(\l_0))
< \wg{g}^2(\wg{f}^{i-2}(\l_0))
< \ldots
< \wg{g}^i(\l_0)
\leq \h^i(\h_0) .$$
At least, the requirements hold if $\WG(t) = \hg(t)$,
thus we can duplicate the result of Barzdin
\cite{Barzdin.Barzdin.1991},
and improve the time complexity for the
optimization of Aiken and Murphy
\cite{Aiken.Murphy.1991b} based on
detecting nonterminals producing the empty language.

In the following section, we present an improved algorithm which takes
at most $\O(al \cdot \log nt)$ time in any case.

\section{Lazy Propagation Algorithm}
\label{Lazy Propagation Algorithm}

As we saw above, a naive transfer of Barzdins' liquid--flow technique to
our framework is not sufficient to obtain a sub--quadratic weight
computation algorithm, except for very special cases.
The reason for this is essentially that the algorithm considers terms
in order of increasing height, cf.\ Lem.~\eqr{42}, but
terms with small height may have large weight and vice versa.
For example, in Fig.~\ref{Algorithm 38 in Exm.},
the value $\yy{Q_1}{2} = 1$ is propagated to
$\yy{P_2}{3} = 2$ and $\yy{Q_2}{3} = 3$;
in the next cycle, the value improves to $\yy{Q_1}{3} = 0$, and both
propagations have to be redone, leading to
$\yy{P_2}{4} = 0$ and $\yy{Q_2}{4} = 1$,
of which the latter is still non--optimal.

The basic idea of the following improved liquid--flow algorithm is
to defer propagation of a $\yy{N}{k}$ value until
we safely know it is the final, minimal one for $N$, following the motto
\notion{less haste more speed}.
The algorithm may need more computation cycles%
	\footnote{%
	In fact, this is not the case, as shown in Lem.~\eqR{49o} below.
	}%
,
but in each cycle less updates are made, and, more important, the right
updates are made.

\ALGORITHM{ \eqd{49k}
\notion{(Lazy propagation)}
~
The weights of all nonterminals can be
computed using the following improved fixpoint algorithm.

Maintain three sets of nonterminals $\FF{k}$, $\MM{k}$, and $\DD{k}$,
called \notion{water front}, \notion{minimals}, and \notion{done},
respectively.
For $N \in \NT$, let
$\yy{N}{0} := \infty$; let
$\FF{0} := \set{}$,
$\MM{0} := \set{}$,
$\DD{0} := \set{}$,
and, for simplicity of proofs,
$\DD{-1} := \set{}$.

For each $N$
and each $k \in \N$, define
$$\begin{array}{@{}l@{\;}l@{}}
\yy{N}{k+1}
	& := \min (\set{ \yy{N}{k}} \cup
	\set{ \wg{f}_i( \,1{n_i}{\yy{{N_{i\i}}}{k}} )
	\mid 1 \leq i \leq m \wedge
	\bigwedge_{j=1}^{n_i} N_{ij} \in \DD{k}}) \\
\FF{k+1}
	& := (\FF{k}
	\cup \set{ N \mid \yy{N}{k+1} < \yy{N}{k}})
	\setminus \DD{k}, \\
\MM{k+1}
	& := \set{N \in \FF{k+1}
	\mid \forall N' \in \FF{k+1}: \;\;
	\yy{N}{k+1} \leq \yy{N'}{k+1}}, \mbox{ ~ ~ and}	\\
\DD{k+1}
	& := \DD{k} \cup \MM{k+1}.	\\
\end{array}$$
Stop if $\FF{k+1} = \set{}$.
\qed
}

Informally, this algorithm works as follows.
When an $\yy{N}{\cdot}$ value changes, i.e.\
$\yy{N}{k+1} < \yy{N}{k}$,
we add $N$ to the \notion{water front} $\FF{k+1}$, as does
Alg.~\eqr{49c}.
However, only those $N$ having minimal $\yy{N}{k+1}$ values in
$\FF{k+1}$ are used for propagation.
They are collected in the set \notion{minimal}, i.e.\ $\MM{k+1}$,
and added to the set \notion{done}, i.e.\ $\DD{k+1}$.

In Lem.~\eqR{49t} below, we will show that their $\yy{N}{\cdot}$ value
in fact doesn't change any further.
This is what is intuitively expected, since only larger values are in
circulation.

At first glance, we observe the following facts about Alg.~\eqr{49k}.

\LEMMA{ \eqd{49q}
\begin{enumerate}
\item $\DD{k} = \bigcup_{i=0}^k \MM{i}$,
	where all $\MM{i}$ are pairwise disjoint.
\item $\DD{k} \subseteq \DD{k+1}$,
	and
	$\DD{k} \subset \DD{k+1}
	\Lra \MM{k+1} \neq \set{}
	\Lra \FF{k+1} \neq \set{}$.
\item $\yy{N}{k} \geq \yy{N}{k+1}$.
\item $N \in \FF{k} \cup \DD{k}$
	iff $\yy{N}{k} < \infty$;
	both implies $\exists k': \; N \in \DD{k'}$.
\item If $N \in \MM{k}$ and $N' \not\in \DD{k-1}$,
	then $\yy{N}{k} \leq \yy{N'}{k}$.
\item $N \in \FF{k+1} \setminus \FF{k}$ iff
	$\yy{N}{k+1} < \infty = \yy{N}{k}$.
\end{enumerate}
}
\pROOF{
\begin{itemize}
\item[1.] -- 3. Follow directly from Alg.~\eqr{49k}.
\item[4.] Show
	$N \not\in \FF{k} \cup \DD{k-1} \Ra \yy{N}{k} = \infty$
	by induction on $k$, using 2.
	\\
	Show $N \in \FF{k} \Ra \yy{N}{k} < \infty$, 
	by induction on $k$, using 3.
	\\
	Show $N \in \DD{k} \Ra \yy{N}{k} < \infty$, using 1.\ and 3.
	\\
	Show
	$N \in \FF{k} \wedge (\forall k': \; N \not\in \DD{k'})
	\Ra \DD{k+nt+1} \not\subseteq \NT$,
	using 2.
\item[5.] Show $N' \in \FF{k} \Ra \yy{N}{k} \leq \yy{N'}{k}$.
	\\
	Show $N' \not\in \FF{k} \Ra \yy{N'}{k} = \infty$,
	using 1.\ and 4.
\item[6.]
\item[7.]
	\qed
\end{itemize}
}
\PROOF{
$\;$
\begin{enumerate}
\item $\DD{k} = \bigcup_{i=0}^k \MM{i}$
        follows from Def.~$\DD{\cdot}$.
        \\
        To see the disjointness,
        note that
	\\
        $N \in \MM{k} \Ra N \in \FF{k} \Ra N \not\in \DD{k-1}$
        by Def.~$\MM{k}, \FF{k}$.
\item $\DD{k} \subseteq \DD{k+1}$
        follows from Def.~$\DD{k+1}$.
        \\
        $\DD{k} \subset \DD{k+1} \Lra \MM{k+1} \neq \set{}$
        follows from 1.
        \\
        $\MM{k+1} \neq \set{} \Lra \FF{k+1} \neq \set{}$
        follows from Def.~$\MM{k+1}$.
\item Follows from Def.~$\yy{N}{k+1}$.
\item First, we show
        $N \not\in \FF{k} \cup \DD{k-1} \Ra \yy{N}{k} = \infty$
        by induction on $k$:
        \\
        $k=0$:
        ~
        $\yy{N}{0} = \infty$ for all $N$.
        \\
        $k \leadsto k+1$:
        \\
        \begin{tabular}[t]{@{}llllll@{\hspace*{1cm}}l@{}}
        &&& $N \not\in \FF{k+1}$ & $\wedge$ & $N \not\in \DD{k}$  \\
        $\Ra$ & $\yy{N}{k+1} = \yy{N}{k}$
                & $\wedge$ & $N \not\in \FF{k}$
                & $\wedge$ & $N \not\in \DD{k-1}$
                & by Def.\ $\FF{\cdot}$ and 2.        \\
        $\Ra$ & \multicolumn{3}{l}{$\yy{N}{k+1} = \yy{N}{k} = \infty$}
                &&& by I.H.        \\
        \end{tabular}
        \\
        Next, if $N \in \FF{k}$,
        let $k'$ be minimal with that property,
	\\
        then $\yy{N}{k} \leq \yy{N}{k'} < \yy{N}{k'-1} \leq \infty$,
        using 3.
        \\
        If $N \in \DD{k}$,
        then $N \in \MM{k'} \subseteq \FF{k'}$
        for some $k' \leq k$ by 1.;
	\\
        hence $\yy{N}{k} \leq \yy{N}{k'} < \infty$ by 3.
        \\
        Finally, assume for contradiction $N \in \FF{k}$
        and $N \not\in \DD{k'}$ for all $k' \in \N$.
        \\
        Then $N \in \FF{k'+1}$ for all $k' \geq k$ by Def.~$\FF{k'+1}$.
        \\
        From 2., we get $\abs{\DD{k+nt+1}} > nt$,
        where $nt = \abs{\NT}$,
	\\
        which contradicts $\DD{k+nt+1} \subseteq \NT$.
\item If $N' \in \FF{k}$,
        we have $\yy{N}{k} \leq \yy{N'}{k}$
        by Def.~$\MM{k}$.
        \\
        If $N' \not\in \FF{k} \supseteq \MM{k}$,
        then $N' \not\in \DD{k}$ by 1.,
        and we have $\yy{N}{k} \leq \infty = \yy{N'}{k}$
        by 4.
\item ``$\Ra$'':
	\\
        \begin{tabular}[t]{@{}ll@{\hspace*{1cm}}l@{}}
	& $N \not\in \FF{k}$ and $N \in \FF{k+1}$	\\
	$\Ra$ & $\yy{N}{k+1} < \yy{N}{k}$ and $N \not\in \DD{k}$
		& Def.~$\FF{k+1}$	\\
	$\Ra$ & $\yy{N}{k} = \infty$
		& by 4.	\\
	\end{tabular}
	\\
	``$\La$'':
	\\
	\begin{tabular}[t]{@{}ll@{\hspace*{1cm}}l@{}}
	& $\yy{N}{k+1} < \infty = \yy{N}{k}$	\\
	$\Ra$ & $N \in \FF{k+1} \vee N \in \DD{k}$
		& Def.~$\FF{k+1}$	\\
	and & $N \not\in \FF{k} \cup \DD{k}$
		& by 4.	\\
	$\Ra$ & $N \not\in \FF{k}$ and $N \in \FF{k+1}$	\\
	\end{tabular}
\end{enumerate}
\qed
}

We now prove the core property of Alg.~\eqr{49k}:
\\
the values corresponding to nonterminals in the \notion{done} set
don't change any more.

\LEMMA{ \eqd{49t}
If $\Na \in \DD{k}$
and $k \leq k'$,
then $\yy{\Na}{k} = \yy{\Na}{k'}$.
}
\PROOF{
$\;$	\\
We show
$$\forall k' \in \N \;\; \forall k \leq k' \;\;
\forall \Na \in \MM{k} \;\; \forall \Nb \not\in \DD{k-1}: \;\;
\yy{\Na}{k} \leq \yy{\Nb}{k'}$$
by induction on $k'$.
The case $k'=0$ is trivial,
since then $k=0$ and $\yy{\Na}{0} = \infty = \yy{\Nb}{0}$.
\\
In the case $k' \leadsto k'+1$,
we have to show
\\
$\yy{\Na}{k} \leq \yy{\Nb}{k'+1}$
for
$k \leq k'+1$ and
$\Na \in \MM{k}, \Nb \not\in \DD{k-1}$.
\\
We are done by Lem.~\eqr{49q}.5 if $k = k'+1$,
so let $k \leq k'$
in the following.
\\
Consider the definition of $\yy{\Nb}{k'+1}$.
\\
If $\yy{\Nb}{k'+1} = \yy{\Nb}{k'}$, we are done
immediately, using the I.H.\
$\yy{\Na}{k} \leq \yy{\Nb}{k'}$.
\\
If
$\yy{\Nb}{k'+1}
= \wg{f}_i(\,1{n_i}{\yy{{\Nb_{i\i}}}{k'}})$
for some $i \in \set{\,1m\i}$
with $\,1{n_i}{\Nb_{i\i}} \in \DD{k'}$,
we distinguish the following cases:
\begin{itemize}
\item $\Nb_{ij} \in \DD{k-1}$ for all $j \in \set{\,1{n_i}\i}$:
	\\
	Note that this implies $k > 0$ or $n_i = 0$.
	\\
	By Lem.~\eqr{49q}.1,
	for each $j=\,1{n_i}\i$ some $k_j \leq k-1$ exists
	with $\Nb_{ij} \in \MM{k_j}$.
	\\
	We have for all $j \in \set{\,1{n_i}\i}$:
	\\
	\begin{tabular}[t]{@{}l@{$\;$}l@{\hspace*{1cm}}l@{}}
	& $\yy{\Nb_{ij}}{k_j}$   \\
	$\leq$ & $\yy{\Nb_{ij}}{k'}$
		& by I.H., since $\Nb_{ij} \in \MM{k_j}$,
		hence $\Nb_{ij} \not\in \DD{k_j-1}$,
		and $k_j \leq k'$       \\
	$\leq$ & $\yy{\Nb_{ij}}{k-1}$
		& by Lem.~\eqr{49q}.3,
		since $k-1 \leq k'$     \\
	$\leq$ & $\yy{\Nb_{ij}}{k_j}$
		& by Lem.~\eqr{49q}.3,
		since $k_j \leq k-1$    \\
	\end{tabular}
	\\
	That is, all these terms are equal;
	in particular, $\yy{\Nb_{ij}}{k-1} = \yy{\Nb_{ij}}{k'}$
	for all $j$.
	Hence:
	\\
	\begin{tabular}[t]{@{}l@{$\;$}l@{\hspace*{1cm}}l@{}}
	& $\yy{\Na}{k}$   \\
	$\leq$ & $\yy{\Nb}{k}$
		& by Lem.~\eqr{49q}.5,
		since $\Na \in \MM{k}, \Nb \not\in \DD{k-1}$       \\
	$\leq$ & $\wg{f}_i(\,1{n_i}{\yy{{\Nb_{i\i}}}{k-1}})$
		& by Alg.~\eqr{49k},
		since $\Nb_{ij} \in \DD{k-1}$
		for all $j$     \\
	$=$ & $\wg{f}_i(\,1{n_i}{\yy{{\Nb_{i\i}}}{k'}})$
                        & as shown above        \\
                $=$ & $\yy{\Nb}{k'+1}$ & by assumption      \\
                \end{tabular}
        \item $\Nb_{ij} \not\in \DD{k-1}$
                for some $j \in \set{\,1{n_i}\i}$:
                \\
                \begin{tabular}[t]{@{}l@{$\;$}l@{\hspace*{1cm}}l@{}}
                & $\yy{\Na}{k}$   \\
                $\leq$ & $\yy{\Nb_{ij}}{k'}$
                        & by I.H., since
			$\Na \in \MM{k}, \;
			\Nb_{ij} \not\in \DD{k-1}, \;
			k \leq k'$ \\
                $\leq$ & $\wg{f}_i(\,1{n_i}{\yy{{\Nb_{i\i}}}{k'}})$
                        & since $\wg{f}_i$ is increasing \\
                $=$ & $\yy{\Nb}{k'+1}$ & by assumption      \\
                \end{tabular}
        \end{itemize}
This completes the induction proof.

Now, if $\Na \in \DD{k}$,
then $\Na \in \MM{k''}$ for some $k'' \leq k$,
\\
and we get
$\yy{\Na}{k''} \leq \yy{\Na}{k'}$ from above
and $\yy{\Na}{k'} \leq \yy{\Na}{k} \leq \yy{\Na}{k''}$
from Lem.~\eqr{49q}.3,
\\
i.e.\ all these values are equal.
\qed
}


\COROLLARY{ \eqd{49v}
Let $N,N' \in \NT$,
~
$k \leq k'$,
~
$N \in \DD{k}$,
~
and $N' \not\in \DD{k-1}$.
~
Then:
\be
\item $\yy{N}{k} = \yy{N}{k'}$, and
\item $\yy{N}{k} \leq \yy{N'}{k'}$.
\ee
}
\PROOF{
$\;$
\be
\item 
        \begin{tabular}[t]{@{}ll@{\hspace*{1cm}}l@{}}
        & $N \in \DD{k}$        \\
        $\Ra$ & $N \in \MM{k''}$ for some $k'' \leq k$
                & by Lem.~\eqr{49q}.1  \\
        $\Ra$ & $N \not\in \DD{k''-1}$
                & by Lem.~\eqr{49q}.1  \\
        $\Ra$ & $\yy{N}{k''} \leq \yy{N}{k'} \leq \yy{N}{k} 
                \leq \yy{N}{k''}$
                & by Lem.~\eqr{49t} and~\eqr{49q}.3,
                since $k'' \leq k \leq k'$      \\
        \end{tabular}
\item
        \begin{tabular}[t]{@{}rl@{\hspace*{1cm}}l@{}}
        & $N \in \DD{k}$        \\
        $\Ra$ & $N \in \MM{k''}$ for some $k'' \leq k$
                & by Lem.~\eqr{49q}.1  \\
        $\wedge$ & $N' \not\in \DD{k''-1}$
                & by Lem.~\eqr{49q}.2,
                since $k''-1 \leq k-1$  \\
        $\Ra$ & $\yy{N}{k} \leq \yy{N}{k''} \leq \yy{N'}{k'}$
                & by Lem.~\eqr{49q}.3 and~\eqr{49t},
                since $k'' \leq k \leq k'$      \\
        \end{tabular}
\ee
\qed
}

\LEMMA{ \eqd{49zz}
$\FF{k+1} = (\FF{k} 
	\cup \set{ N \mid \yy{N}{k+1} < \yy{N}{k} }) \setminus \MM{k}$.
}
\PROOF{
``$\subseteq$'':
~
obvious, since $\DD{k} \supseteq \MM{k}$ by \ref{49q}.1.
\\
``$\supseteq$'':
~
If $N \in \FF{k}$ and $N \not \in \MM{k}$,
then
$N \not\in \DD{k-1}$ by Def.~$\FF{k}$,
\\
hence $N \not\in \DD{k} = \DD{k-1} \cup \MM{k}$ by \ref{49q}.1.
\\
If $\yy{N}{k+1} < \yy{N}{k}$ and $N \not \in \MM{k}$,
\\
then $N \not\in \DD{k}$,
since else $\yy{N}{k+1} = \yy{N}{k}$ by Lem.~\ref{49t}.
\qed
}

\LEMMA{ \eqd{49m}
$N \in \DD{k} \Ra \exists t \in \mL(N): \;\; \WG(t) = \yy{N}{k}$.
}
\pROOF{
Induction on $k$:
If $k' \leq k+1$ such that $N \in \MM{k'} \subseteq \DD{k+1}$,
show that
$$\begin{array}{@{}r@{\;}l@{}}
k''' := \max \set{k''
	\mid
	& k'' \leq k' \; \wedge	\\
	& \exists i: \;
	\yy{N}{k''} = \wg{f}_i(\,1{n_i}{\yy{N_{i\i}}{k''-1}})
	\wedge
	\forall j: \; N_{ij} \in \DD{k''-1}}
\end{array}$$
is well--defined,
show $\yy{N}{k+1} = \wg{f}_i(\,1{n_i}{\yy{N_{i\i}}{k'''-1}}) =
\WG(f_i(\,1{n_i}{t_\i}))$ for some $i$, $\,1{n_i}{t_\i}$,
using Lem.~\eqr{49t}.
}
\PROOF{
Induction on $k$:
\begin{itemize}
\item $k = 0$: trivial, since $\DD{0} = \set{}$.
\item $k \leadsto k+1$:
        \\
	First, we have
	\\
        \begin{tabular}[t]{@{}ll@{\hspace*{1cm}}l@{}}
        & $N \in \DD{k+1}$      \\
        $\Ra$ & $N \in \MM{k'} \subseteq \FF{k'}$
                for some $k' \leq k+1$
                & by Lem.~\eqr{49q}.1  \\
        $\Ra$ & $\yy{N}{k''} < \yy{N}{k''-1}$ 
                for some $k'' \leq k'$
                & by Def.~$\FF{\cdot}$  \\
        $\Ra$ & $\yy{N}{k''} 
		= \wg{f}_i(\,1{n_i}{\yy{N_{i\i}}{k''-1}})$	\\
	& for some $i$
                with $\forall j: N_{ij} \in \DD{k''-1}$
                & by Alg.~\eqr{49k}     \\
        \end{tabular} 
        \\
        Let $k'''$ be the maximal $k''$ with that property,
        i.e., let
        \\
        $k''' \.= \max \set{k'' \.\in \N
                \mid k'' \.\leq k', \;
                \exists i: 
                \yy{N}{k''} 
		\.= \wg{f}_i(\,1{n_i}{\yy{N_{i\i}}{k''\.-1}})
                \wedge
                \forall j: N_{ij} \.\in \DD{k''\.-1}}$.
        \\
        Then:
        \\
        \begin{tabular}[b]{@{}ll@{\hspace*{1cm}}l@{}}
        & $\yy{N}{k+1}$ \\
        $=$ & $\yy{N}{k'}$
                & by Cor.~\eqr{49v}.1,
                since $N \in \DD{k'}$   \\
        $=$ & $\:={k'-1}{k'''}{\yy{N}{\i}}$
                & by Def.~$k'''$        \\
        $=$ & $\wg{f}_i(\,1{n_i}{\yy{N_{i\i}}{k'''-1}})$	\\
	& for some $i$
                with $\forall j: N_{ij} \in \DD{k'''-1}$
                & by Def.~$k'''$        \\
        $=$ & $\wg{f}_i(\,1{n_i}{\WG(t_\i)})$	\\
	& for some $t_j \in \mL(N_{ij})$
                & by I.H., since $k''' \leq k' \leq k+1$        \\
        $=$ & $\WG(f_i(\,1{n_i}{t_\i}))$
                & by Def.~$\WG(\cdot)$  \\
        \end{tabular}
	\qed
\end{itemize}
}

\LEMMA{ \eqd{29a}
Let $N \in \NT$ with $\mL(N) \neq \set{}$ be given.
\\
Then, a derivation $N \stackrel{*}{\raa} t$
exists such that each
subterm $t'$ of $t$ is minimal wrt.\ the nonterminal
$\nu(t')$ it has been derived from
~
(Note that in a nondeterministic grammar,
$\nu(\cdot)$ is well--defined only in the context of a given
derivation).
\\
We call such a $t$ \notion{thoroughly minimal\/}.
}
\PROOF{
$\;$	\\
We show that for each derivation of a minimal $t$ wrt.\ $N$
a derivation of some thoroughly minimal $t''$ wrt.\ $N$ exists,
by induction on the structure of $t$:
\\
Let $N \raa f(\,1n{N_\i}) \stackrel{*}{\raa} f(\,1n{t_\i}) = t$
and $t$ be minimal wrt.\ $N$.
\\
Since $t_i \.\in \mL(N_i) \.\neq \set{}$, 
we can find a derivation
$N_i \stackrel{*}{\raa} t'_i$ of some minimal $t'_i$ wrt.\ $N_i$.
\\
By I.H., we find for $i=\,1n\i$ a derivation
$N_i \stackrel{*}{\raa} t''_i$ of some thoroughly minimal $t''_i$ wrt.\
$N_i$.
\\
Let $t'' := f(\,1n{t''_\i})$,
\\
then
$\WG(t'')
= \wg{f}(\,1n{\WG(N_\i)})
\leq \wg{f}(\,1n{\WG(t_\i)})
= \WG(t)
= \WG(N)$,
\\
hence $t''$ is minimal, and therefor also thoroughly minimal, wrt.\ $N$.
\qed
}

\LEMMA{ \eqd{49x}
If $\WG(N) < \infty$,
then $N \in \DD{k}$ for some $k$.
}
\PROOF{
$\;$	\\
If $\WG(N) < \infty$,
we can find some derivation $N \stackrel{*}{\raa} t$
of some thoroughly minimal $t$ wrt.\ $N$ by Lem.~\eqr{29a}.
\\
We show that $t$ thoroughly minimal wrt.\ $N$ and $\WG(t) < \infty$
implies $\yy{N}{k} \leq \WG(t)$ and $N \in \DD{k}$ for some $k \in \N$,
by induction on the structure of $t$:
\\
Let $t = f(\,1n{t_\i})$.
~
Then $N \raa f(\,1n{N_\i})$ for some $N_i$,
each $t_i$ is thoroughly minimal wrt.\ $N_i$,
and $\WG(N_i) = \WG(t_i) \leq \WG(t) < \infty$.
\\
By I.H., we get some $k_i$ with $\yy{N_i}{k_i} \leq \WG(t_i)$ and
$N_i \in \DD{k_i}$ for $i=\,1n\i$.
\\
Let $k := \max \set{ \,1n{k_\i} }$.
~
Then
\\ 
\begin{tabular}[t]{@{}ll@{\hspace*{1cm}}l@{}}
& $\yy{N}{k+1}$ \\
$\leq$ & $\wg{f}(\,1n{\yy{N_\i}{k}})$
        & by Alg.~\eqr{49k},
        since $N_i \in \DD{k}$ by Lem.~\eqr{49q}.2     \\
$\leq$ & $\wg{f}(\,1n{\WG(t_\i)})$
        & since $\yy{N_i}{k} \leq \yy{N_i}{k_i} \leq \WG(t_i)$
        by Lem.~\eqr{49q}.3    \\
$=$ & $\WG(t)$ & by Def.~$\WG(\cdot)$   \\
$<$ & $\infty$ & by assumption  \\
\end{tabular}
\\
From Lem.~\eqr{49q}.4, we get $N \in \DD{k'}$ for some $k'$.
\\
Setting $k'' := \max \set{k+1,k'}$,
\\
we have
$\yy{N}{k''} \leq \yy{N}{k+1} \leq \WG(t)$
and $N \in \DD{k'} \subseteq \DD{k''}$
by Lem.~\eqr{49q}.3 and 2.
\qed
}

\LEMMA{ \eqd{49l}
$\FF{k+1} = \set{}$
iff $\DD{k} = \set{ N \in \NT \mid \WG(N) < \infty}$.
\\
In this case, we reached a fixpoint,
\\
i.e.\
$\MM{k+1} = \set{}$,
$\DD{k+1} = \DD{k}$,
$\yy{N}{k+2} = \yy{N}{k+1}$,
and $\FF{k+2} = \set{}$.
}
\pROOF{
To prove the fixpoint property, show
$\DD{k} = \DD{k-1} \Ra \yy{N}{k+1} \geq \yy{N}{k}$
using Lem.~\eqr{49t}.
To prove the equivalence:
\begin{itemize}
\item Observe that
	$N \in \FF{k+1} \wedge \WG(N) = \infty$
	contradicts
	Lem.~\eqr{49m}
	and \eqr{49q}.2.
\item Show
	$N \in \DD{k} \Ra \WG(N) < \infty$,
	using Lem.~\eqr{49m}
	and \eqr{49q}.4.
\item Show
	$\WG(N) < \infty \Ra \exists k': \; N \in \DD{k'}$,
	by assuming a derivation of some $t$ from $N$ such that each
	subterm of $t$ is minimal wrt.\ the nonterminal it has been
	derived from%
		\footnote{%
		Such a derivation can be constructed by successively
		``correcting'' an arbitrary derivation of some
		$t' \in \mL(N)$.
		}%
	,
	and showing that such a $t$ with $\WG(t) < \infty$
	always satisfies
	$\yy{N}{k'} \leq \WG(t)$ and $N \in \DD{k'}$ 
	for some $k' \in \N$,
	by induction on the structure of $t$.
\item Observe that this $k'$ can be assumed to equal $k$,
	if $\FF{k+1} = \set{}$.
	\qed
\end{itemize}
}
\PROOF{
$\;$
\begin{enumerate}
\item $\DD{k+1} = \DD{k} \Ra \yy{N}{k+2} = \yy{N}{k+1}$:
	\\
        \begin{tabular}[t]{@{}ll@{}l@{}ll@{\hspace*{0.8cm}}l@{}}
        & $\yy{N}{k+2}$ \\
        $=$ & $\min (\set{ \yy{N}{k\.+1}}$
                & $\cup$
                & $\set{ \wg{f}_i( ...\yy{N_{ij}}{k+1}... )$
                & $\mid 1 \.\leq i \.\leq m, \;
                \bigwedge_{j=1}^{n_i} N_{ij} \.\in \DD{k\.+1}})$
                & Alg.\eqr{49k}	\\
        $=$ & $\min (\set{ \yy{N}{k\.+1}}$
                & $\cup$
                & $\set{ \wg{f}_i( ...\yy{N_{ij}}{k+1}... )$
                & $\mid 1 \.\leq i \.\leq m, \;
                \bigwedge_{j=1}^{n_i} N_{ij} \.\in \DD{k}})$
                & ass.	\\
        $=$ & $\min (\set{ \yy{N}{k\.+1}}$
                & $\cup$
                & $\set{ \wg{f}_i( ...\yy{N_{ij}}{k}... )$
                & $\mid 1 \.\leq i \.\leq m, \;
                \bigwedge_{j=1}^{n_i} N_{ij} \.\in \DD{k}})$
                & Cor.\eqr{49v}.1	\\
        $\geq$ & $\min (\set{ \yy{N}{k\.+1}, \yy{N}{k}}$
                & $\cup$
                & $\set{ \wg{f}_i( ...\yy{N_{ij}}{k}... )$
                & $\mid 1 \.\leq i \.\leq m, \;
                \bigwedge_{j=1}^{n_i} N_{ij} \.\in \DD{k}})$
                & $\min$	\\
        $=$ & $\yy{N}{k+1}$ &&&& Alg.\eqr{49k}	\\
        $\geq$ & $\yy{N}{k+2}$ &&&& Lem.\eqr{49q}.3	\\
        \end{tabular}
\item If $\FF{k+1} = \set{}$, we reached a fixpoint:
	\\
        \begin{tabular}[t]{@{}lrl@{\hspace*{1cm}}l@{}}
        & $\FF{k+1}$ & $= \set{}$ \\
        $\Ra$ & $\MM{k+1}$ & $= \set{}$
                & by Def.\ $\MM{k+1}$ \\
        $\Ra$ & $\DD{k+1}$ & $= \DD{k}$
                & by Lem.~\eqr{49q}.1  \\
        $\Ra$ & $\yy{N}{k+2}$ & $= \yy{N}{k+1}$ for all $N \in \NT$
                & by 1. \\
        $\Ra$ & $\FF{k+2}$ & $= \set{}$
                & by Def.\ $\FF{\cdot}$,
                since $\FF{k+1} = \set{}$ \\
        \end{tabular}
\item $\FF{k+1} = \set{} 
	\La \DD{k} = \set{ N \in \NT \mid \WG(N) < \infty}$:
        \\
        Assume for contradiction $\FF{k+1} \neq \set{}$.
        \\
        By Lem.~\eqr{49q}.2,
        we have $N \in \MM{k+1} \subseteq \FF{k+1}$
        for some $N$ with $\WG(N) = \infty$.
        \\
        By Lem.~\eqr{49m},
        $\WG(t) = \yy{N}{k+1}$ for some $t \in \mL(N)$.
        \\
        Hence,
        $\WG(N) \leq \WG(t) = \yy{N}{k+1} < \infty$,
        which is a contradiction.
\item Let $\FF{k+1} = \set{}$,
        we show $\WG(N) < \infty \Lra N \in \DD{k}$:
        \\
        \begin{tabular}[t]{@{}rl@{\hspace*{1cm}}l@{}}
        & $\WG(N) < \infty$     \\
        $\Ra$ & $N \in \DD{k'}$ for some $k'$
                & by Lem.~\eqr{49x}    \\
        $\Ra$ & $N \in \DD{k}$
                & since $\DD{k'} \subseteq \DD{k}$
                by 2.\ and Lem.~\eqr{49q}.2    \\
        $\Ra$ & $\WG(N) \leq \yy{N}{k}$
                & by Lem.~\eqr{49m}    \\
        $\wedge$ & $\yy{N}{k} < \infty$
                & by Lem.~\eqr{49q}.4  \\
        \end{tabular}
\end{enumerate}
\qed
}

\LEMMA{ \eqd{49n}
If $\FF{k+1} = \set{}$,
then $\yy{N}{k} \leq \WG(t)$
for all $N \in \NT$ and all $t \in \mL(N)$.
}
\pROOF{
Induction on $t$, using Lem.~\eqr{49l}.
}
\PROOF{
Induction on $t$:
\\
Let $t = f( \,1n{t_\i} ) \in \mL(N)$,
\\
i.e., $N \sortdef \ldots f(\,1n{N_\i}) \ldots$
and $t_j \in \mL(N_j)$ for $j=\,1n\i$.
\\
If $\WG(t_j) = \infty$ for some $j$,
then $\yy{N}{k} \leq \infty = \WG(t)$,
since $\wg{f}$ is increasing.
\\
If $\WG(t_j) < \infty$ for all $j$,
then
\\
\begin{tabular}[t]{@{}ll@{\hspace*{1cm}}l@{}}
& $\yy{N}{k}$   \\
$=$ & $\yy{N}{k+1}$ & by Lem.~\eqr{49l}.2      \\
$\leq$ & $\wg{f}( \,1n{\yy{N_\i}{k}} )$
        & by Alg.~\eqr{49k},
        since $N_j \in \DD{k}$ for all $j$
        by Lem.~\eqr{49l}.3    \\
$\leq$ & $\wg{f}( \,1n{\WG(t_\i)} )$
        & I.H., $\wg{f}$ monotonic      \\
$=$ & $\WG(t)$ & Def.~$\WG$     \\
\end{tabular}
\\
\qed
}

\COROLLARY{ \eqd{49o}
\notion{(Correctness of Alg.~\eqr{49k})}
~
If $\FF{k+1} = \set{}$,
then $\yy{N}{k} = \WG(N)$ for all $N \in \NT$.
}
\pROOF{
We have
$\yy{N}{k} \leq \WG(N)$
by Lem.~\eqr{49n}.
If $N \in \DD{k}$,
we have $\yy{N}{k} \geq \WG(N)$
by Lem.~\eqr{49m};
else, we have
$\yy{N}{k} = \infty = \WG(N)$ by Lem.~\eqr{49q}.4 and \eqr{49l}.
}
\PROOF{
$\;$	\\
By Lem.~\eqr{49q}.2, we have $\FF{k+1} = \set{}$
for some $k \leq nt$.
\\
By Lem.~\eqr{49l}.2 and 3, thus $\FF{nt+1} = \set{}$
and $\DD{nt} = \set{ N \in \NT \mid \WG(N) < \infty}$.
\\
By Lem.~\eqr{49n}, we have $\yy{N}{nt} \leq \WG(N)$
for all $N \in \NT$.
\\
By Lem.~\eqr{49m}, we have $\yy{N}{nt} \geq \WG(N)$ for
all $N \in \DD{nt}$.
\\
For all $N \in \NT \setminus \DD{nt}$,
we have
$\forall k: \; N \not\in \DD{k}$ by Lem.~\eqr{49l}.2 and \eqr{49q}.2,
\\
hence $\WG(N) = \infty = \yy{N}{nt}$ by Lem.~\eqr{49q}.4.
\qed
}

We now duplicate the transformation from Alg.~\ref{38} 
to Alg.~\ref{49c} for Alg.~\ref{49k}:

\ALGORITHM{ \label{50a}
Replace in Alg.~\ref{49k}
the definition rule for $\yy{N}{k+1}$, and $\FF{k+1}$ by
$$\begin{array}{@{}l@{\;}l@{\;}ll@{}}
\yy{N}{k+1} & :=
	\min (\set{ \yy{N}{k}} \cup
	\{ 
	& \wg{f}_i( \,1{n_i}{\yy{{N_{i\i}}}{k}} )
	\mid
	1 \leq i \leq m  \; \wedge \\
&& (\bigwedge_{j=1}^{n_i} N_{ij} \in \DD{k})
	\wedge (\bigvee_{j=1}^{n_i} N_{ij} \in \MM{k}) \})
	& \mbox{, and} \\
\FF{k+1} & \multicolumn{2}{l}{:= (\FF{k}
        \cup \set{ N \mid \yy{N}{k+1} < \yy{N}{k}})
        \setminus \MM{k}} & , \\
\end{array}$$
while all other definitions remain unchanged.
\qed
}

\LEMMA{
\notion{(Correctness of Alg.~\ref{50a})}
~
In Alg.~\ref{50a}, each $\yy{N}{k}$, and each $\FF{k}$
has the same value in Alg.~\ref{50a}
as in Alg.~\ref{49k}.
}
\PROOF{
``$\yy{N}{k}$'':
~
Induction on $k$ using Lem.~\eqr{49t}.
\\
``$\FF{k}$'':
~
follows immediately from Lem.~\ref{49zz}.
\qed
}

To estimate the complexity of Alg.~\ref{50a}, we need the
following

\LEMMA{ \label{50b}
For each alternative
$\wg{f}_i( \,1{n_i}{\yy{{N_{i\i}}}{k}} )$,
\\
the formula
$(\bigwedge_{j=1}^{n_i} N_{ij} \in \DD{k})
\wedge (\bigvee_{j=1}^{n_i} N_{ij} \in \MM{k})$
holds for at most one $k$.
}
\PROOF{
For each $j \in \set{\,1{n_i}\i}$,
$N_{ij}$ is in at most one $\MM{k}$ by Lem.~\eqr{49q}.1.
\\
Denoting this $k$ by $k_j$,
we have $N_{ij} \not\in \DD{k}$ for all $k < k_j$,
again by Lem.~\eqr{49q}.1.
\\
Hence, the formula can hold at most for $k = \max \set{\,1{n_i}{k_\i}}$.
\qed
}

\COROLLARY{ \label{50c}
\notion{(Complexity of Alg.~\ref{50a})}
\\
Algorithm~\ref{50a} has a time complexity of
$\O(al \cdot (ar + \log nt))$.
\\
Its memory complexity is that of the input grammar, viz.\
$\O(al \cdot ar)$.
}
\PROOF{
Due to the different access modes,
the sets $\FF{\cdot}$, $\MM{\cdot}$, and $\DD{\cdot}$
can be implemented by a heap
\cite[Sect.\ 3.4]{Aho.Hopcroft.Ullman.1974},
a linked list, and a bit vector,
respectively.
~
Due to Lem.~\ref{49q}.6, it is sufficient to enter a nonterminal $N$
into the heap iff $\yy{N}{\cdot}$ is decreased for the first time.

We add a counter in the range $\set{\,0{ar}\i}$ to each alternative
$$N \sortdef \ldots f_i( \,1{n_i}{N_{i\i}} ) \ldots ,$$
which counts the number of $N_{ij}$ with $N_{ij} \in \DD{k}$.
~
It is initially $0$ and is increased when some $N_{ij}$ is added to
$\DD{k}$ and we visit all alternatives $N_{ij}$ occurs in.
~
If the counter of an alternative
reaches its arity $n_i$, we know that the formula
$$(\bigwedge_{j=1}^{n_i} N_{ij} \in \DD{k})
\wedge (\bigvee_{j=1}^{n_i} N_{ij} \in \MM{k})$$
holds, and evaluate the alternative, setting
$\yy{N}{k+1}
:= \min \set{\yy{N}{k}, \; \wg{f}_i( \,1{n_i}{\yy{{N_{i\i}}}{k}} )}$.
~
It may be neccessary to reorder the heap $\FF{\cdot}$ to get $N$ to its
appropriate place.
~
Since the value of $\yy{N}{\cdot}$ cannot have grown,
it is sufficient to move $N$ upwards in the heap.
~
After evaluation, the alternative
will not be considered any more during the algorithm.

For a certain alternative,
increasing and checking the counter takes at most
$ar \cdot \O(1)$ time,
evaluation of $\wg{f}$ takes $\O(ar)$ time,
reordering the heap takes $\O(\log nt)$ time,
and all other set operations are dominated by the latter.
~
Hence, the overall time complexity is
$\O(al \cdot (ar + \log nt))$.
~
The memory complexity is obvious, since the input grammar dominates
all other data structures, including the counters.

If we omit the counters, we get a time complexity of
$\O(al \cdot (ar^2 + \log nt))$,
since each of the $n_i$ times
we visit an alternative $N_{ij}$ occurs in, we have to
test whether $\bigwedge_{j'=1}^{n_i} N_{ij'} \in \DD{k}$.
~
For small values of $ar$, this complexity may be acceptable,
and we can avoid to extend a possibly huge input grammar by additional
counter fields.
~
If alternatives are stored in prefix form in memory, i.e.\ one word for
$f$, followed by $n_i$ words for $\,1{n_i}{N_{i\i}}$,
it is sufficient to add two binary flags to each nonterminal of each
alternative, one indicating whether it has been visited, the other
indicating whether it is the rightmost argument of its alternative.
~
The latter flag should be set in the word for $f$, too, in order to
delimit the memory area for $\,1{n_i}{N_{i\i}}$ to
the left.
\qed
}

\begin{figure}
\begin{center}
\begin{tabular}[t]{@{}|r||*{5}{r@{$\;$}c@{$\;$}c@{$\;$}c|}@{}}
\hline
$k$
	& \multicolumn{4}{c|}{$Q_0$}
	& \multicolumn{4}{c|}{$P_1$}
	& \multicolumn{4}{c|}{$Q_1$}
	& \multicolumn{4}{c|}{$P_2$}
	& \multicolumn{4}{c|}{$Q_2$}	\\
$\;$
	&& $\yY$ & $\Ff$ & $\Dd$
	&& $\yY$ & $\Ff$ & $\Dd$
	&& $\yY$ & $\Ff$ & $\Dd$
	&& $\yY$ & $\Ff$ & $\Dd$
	&& $\yY$ & $\Ff$ & $\Dd$	\\
\hline
\hline
$0$
	&& $\infty$ &&
	&& $\infty$ &&
	&& $\infty$ &&
	&& $\infty$ &&
	&& $\infty$ &&	\\
\hline
$1$
	& $\g$ & $0$ & $+$ & $+$
	&&&&
	&&&&
	&&&&
	&&&&	\\
\hline
$2$
	&&&& $+$
	& $\p\g$ & $0$ & $+$ & $+$
	& $\j\g$ & $1$ & $+$ &
	&&&&
	&&&&	\\
\hline
$3$
	&&&& $+$
	&&&& $+$
	& $\q\p\g$ & $0$ & $+$ & $+$
	&&&&
	&&&&	\\
\hline
$4$
	&&&& $+$
	&&&& $+$
	&&&& $+$
	& $\p\q\p\g$ & $0$ & $+$ & $+$
	& $\j\q\p\g$ & $1$ & $+$ &	\\
\hline
$5$
	&&&& $+$
	&&&& $+$
	&&&& $+$
	&&&& $+$
	& $\q\p\q\p\g$ & $0$ & $+$ & $+$	\\
\hline
$6$
	&&&& $+$
	&&&& $+$
	&&&& $+$
	&&&& $+$
	&&&& $+$	\\
\hline
\end{tabular}
\caption{Lazy propagation algorithm Alg.~\eqr{49k}}
\label{Lazy propagation algorithm in Exm.}
\end{center}
\end{figure}

Consider again the grammar and the weight functions from
Fig.~\ref{Grammar and weight functions in Exm.}.
Figure~\ref{Lazy propagation algorithm in Exm.}
shows the cycles of Alg.~\eqr{49k} in computing the weights of
each $Q_n$ and $P_n$, where $n_{\max} = 2$.
For each nonterminal $N$, we show (from left to right)
a term of weight $\yy{N}{k}$, the
value of $\yy{N}{k}$, a flag indicating whether $N \in \FF{k}$, and flag
indicating whether $N \in \DD{k}$.
A ``$+$'' flag indicates that the corresponding relation holds, an empty
flag field indicates the contrary.
An empty $\yY$ field means that $\yy{N}{k} = \yy{N}{k-1}$.
The algorithms stops after cycle $6$ since $\FF{6} = \set{}$.
Note that each alternative of each rule is evaluated just once.

\section{Partial weight orderings}
\label{Partial weight orderings}

It would be desirable to provide an equivalent to Alg.~\ref{50a}
for a partial weight order $(<)$ on $\D$.
This would allow us to define weight functions such that $\WG(t)$ is the
set of distinct variables occuring in $t$.
In many applications, a term is more interesting if it contains fewer
distinct variables.
For example, if terms denote transformation rules,
$f(x) \leadsto g(x)$ is usually preferred over
$f(x) \leadsto g(y)$.
However, we have the following negative results.

\LEMMA{ \eqd{51}
There is no total nontrivial
order $(<)$ on the power set $\wp({\cal V})$ of variables
such that $V \subseteq W \Ra V \leq W$ for all $V,W \subseteq {\cal V}$
and $(\cup)$ is monotonic wrt.\ $(<)$.
}
\PROOF{
Let $x,y,z \in {\cal V}$.
If $\set{x,y} < \set{x,z}$,
then $\set{y} \leq \set{x,y} < \set{x,z}$
and trivially $\set{x,z} \leq \set{x,z}$;
hence $\set{x,y,z} \leq \set{x,z}$
by monotonicity.
Similarly, $\set{x,z} < \set{x,y}$ implies $\set{x,y,z} \leq \set{x,y}$.
In each case, there are two sets $V \neq W$ such that $V \leq W \leq V$.
\qed
}


\DEFINITION{ \eqd{52}
It is straight-forward to generalize the definitions from
Sect.~\ref{Grammars and weights} to non-total orderings.
A partial order $(<)$ on the power set $\wp({\cal V})$ of variables
can be defined by: $v_1 < v_2 \Lra v_1 \subsetneq v_2$ for all $v_1, v_2
\subseteq {\cal V}$.

If we define 
$\wg{x} = \set{x}$
for each variable $x \in {\cal V}$,
and
$\wg{f}(\,1n{v_\i}) = \:{\cup}1n{v_\i}$ 
for each $n$-ary function, including constants,
we get $\WG(t) = \var(t)$.
That is, a weight is a set of variables.

Not every set of weights has a unique minimal element,
since $(<)$ is not total.
Therefore, we define for a set $T$ of terms
$$\WG(T) = \set{ v \subseteq {\cal V} \mid 
(\exists t \in T: \; \WG(t) = v)
\land (\lnot \exists t' \in T: \; \WG(t) < v) } .$$
In contrast to Sect.~\ref{Grammars and weights},
$\WG(T) \subseteq \D$ is a set of weights rather than a single weight.
\qed
}

\newcommand{\COMMENT}[1]{{\bf \{COMMENT\}}}

The following notion is needed for Lem.~\eqr{53} below.

\newsavebox{\bigcupPbox}
\savebox{\bigcupPbox}{$\bigcup\hspace{-0.73em}\cdot\hspace{0.30em}$}
\newcommand{\bigcupP}{\usebox{\bigcupPbox}}

\DEFINITION{ \eqd{54}
For $m$ sets $\,1m{S_\i}$ of sets,
define
$\bigcupP_{i=1}^m S_i 
= \set{ \bigcup_{i=1}^m s_i 
\mid \bigwedge_{i=1}^m s_i \in S_i }$
as their pointwise union.

For example, if $S_1 = \set{ \set{a,b,c}, \set{a,d,e} }$
and $S_2 = \set{ \set{b,c,d}, \set{d,e} }$,
we have
$\bigcupP_{i=1}^2 S_i
= \set{ \set{a,b,c,d}, \set{a,b,c,d,e}, \set{a,d,e} }$.
\qed
}

\LEMMA{ \eqd{55}
The pointwise set union from Def.~\eqr{54} has the following properties:
\begin{enumerate}
\item If ~ $\bigwedge_{i=1}^m s \in S_i$,
	~ then ~ $s \in \bigcupP_{i=1}^m S_i$.
\item If ~ $s \in \bigcupP_{i=1}^m S_i$,
	~ then ~ 
	$\bigwedge_{i=1}^m \exists s_i \in S_i: s_i \subseteq s$.
\item If ~ $\abs{s} = n$
	~ and ~
	$\bigwedge_{i=1}^m \bigwedge_{s_i \in S_i} \abs{s_i} = n$,
	~ then ~ 
	$\bigwedge_{i=1}^m s \in S_i \Lra s \in \bigcupP_{i=1}^m S_i$.
\end{enumerate}
}
\PROOF{
$\;$
\begin{enumerate}
\item By Def.~\eqr{54},
	$s = \bigcup_{i=1}^m s$ is an element of $\bigcupP_{i=1}^m S_i$.
\item If $s = \bigcup_{i=1}^m s_i$,
	then $s_i \subseteq s$ for each $i$.
\item Follows from~1 and~2.
\qed
\end{enumerate}
}

\LEMMA{ \eqd{53}
The problem to compute minimal nonterminal weights wrt.\ $(<)$ from
Def.~\eqr{52} is NP hard.
}
\PROOF{
Let a conjunctive normal form $C$ be given.
Let $\set{ \,1n{x_\i} }$ be the set of propositional variables
occurring in $C$.
Without loss of generality, we assume
$$\begin{array}{ll@{\hspace{1cm}}l}
C & \Lra \:{\land}1m{D_\i} & \mbox{ and}	\\
D_i & \Lra \:{\lor}1{n_i}{A_{i\i}} 
	& \mbox{ for } i=\,1{n_i}\i \mbox{, where}	\\
A_{ij} & \in \set{ x_j, (\lnot x_j) } 
	& \mbox{ for } i=\,1m\i 
	\mbox{ and } j=\,1{n_i}\i \mbox{ .}	\\
\end{array}$$
We use a signature
$\Sigma 
= \set{ c, d } \cup \set{ \,1n{y_\i}} \cup \set{ \,1n{z_\i} }$
and consider $\,1n{y_\i}$ and $\,1n{z_\i}$ as variables.
Consider the following grammar:
$$\begin{array}[t]{@{}lll@{\hspace{1cm}}l@{}}
C'   & \sortdef & c( \,1m{D'_\i} ).	\\
D'_i & \sortdef & \:{\mid}1{n_i}{\psi(A_{i\i})}, 
	& i = \,1m\i.	\\
P_j & \sortdef & y_j, & j = \,1n\i.	\\
N_j & \sortdef & z_j, & j = \,1n\i.	\\
F_j & \sortdef & y_j \mid z_j, & j = \,1n\i,	\\
\end{array}$$
where
$\psi$ is a mapping from atoms to alternatives defined by:
$$\begin{array}[t]{@{}*{3}{l}@{}}
\psi(x_j) & = & d(F_1, \ldots , P_j , \ldots , F_n)	\\
\psi(\lnot x_j) & = & d(F_1, \ldots , N_j , \ldots , F_n)	\\
\end{array}$$

The grammar has 
$1 + m + n \cdot 3$ rules and
$1+(\sum_{i=1}^m n_i) + n \cdot 4$ alternatives,
while the conjunctive normal form has $\sum_{i=1}^m n_i$ atoms.
Observe that the set $\L(C')$ 
of all terms derivable from $C'$ is finite,
since the grammar does not contain recursive rules;
similarly, all $\L(D'_i)$ are finite.
If $\alpha$ is a nonterminal or an alternative,
define $\xi(\alpha) = \set{ \var(t) \mid t \in \L(\alpha) }$.

For example, from 
$(x_1 \lor \lnot x_3)
\land (\lnot x_2 \lor x_3)$,
we obtain the grammar
$$\begin{array}{@{}l@{}}
\begin{array}[t]{@{}lll@{}}
C'   & \sortdef & c(D'_1,D'_2),	\\
D'_1 & \sortdef & d(P_1,F_2,F_3) \mid d(F_1,F_2,N_3),	\\
D'_2 & \sortdef & d(F_1,N_2,F_3) \mid d(F_1,F_2,P_3),	\\
\end{array}
\\
\begin{array}[t]{@{}lll*{2}{@{\hspace{0.5cm}}lll}@{}}
P_1 & \sortdef & y_1,	&
P_2 & \sortdef & y_2,	&
P_3 & \sortdef & y_3,	\\
N_1 & \sortdef & z_1,	&
N_2 & \sortdef & z_2,	&
N_3 & \sortdef & z_3,	\\
F_1 & \sortdef & y_1 \mid z_1,	&
F_2 & \sortdef & y_2 \mid z_2,	&
F_3 & \sortdef & y_3 \mid z_3.	\\
\end{array}
\\
\end{array}$$

We call a mapping 
$\sigma: \set{ \,1n{x_\i} } \raa \set{ {\it true}, {\it false} }$
a truth value assignment.
Each such mapping can be homomorphically extended to all propositional
formulas over $\set{ \,1n{x_\i} }$.

Define 
$V = \set{ \set{ \,1n{w_\i} } 
\mid \bigwedge_{j=1}^n w_j \in \set{ y_j, z_j} }$.
Each member of $V$ is a set of cardinality $n$.
Define a bijective mapping
$\rho$ from truth value assignments to $V$ by
$\rho(\sigma) 
= \set{ y_j \mid \sigma(x_j) = {\it true} }
\cup \set{ z_j \mid \sigma(x_j) = {\it false} }$.
%


Observe the following facts:
\begin{enumerate}
\item \label{53 card atoms}
	Each $\xi(\psi(A_{ij}))$ is a subset of $V$:

	This follows from the definition of $\psi(A_{ij})$.
	In the example,
	$\xi(\psi(\lnot x_3))
	= \set{
	\set{y_1,y_2,z_3},\allowbreak
	\set{y_1,z_2,z_3},\allowbreak
	\set{z_1,y_2,z_3},\allowbreak
	\set{z_1,z_2,z_3} }$.

\item \label{53 card disjuncts}
	Each $\xi(D'_i)$ is a subset of $V$:

	This follows from \ref{53 card atoms},
	using
	$\L(D'_i) = \bigcup_{j=1}^{n_i} \L(\psi(A_{ij}))$
	and the definition of $\xi$.
	In the example, we have
	$$\begin{array}{@{}l@{}}
	\xi(D'_1)
	= \set{
	\set{y_1,y_2,y_3},\allowbreak
	\set{y_1,y_2,z_3},\allowbreak
	\set{y_1,z_2,y_3},\allowbreak
	\set{y_1,z_2,z_3},\allowbreak
	\set{z_1,y_2,z_3},\allowbreak
	\set{z_1,z_2,z_3} },	\\
	\xi(D'_2) = \set{
	\set{y_1,y_2,y_3},\allowbreak
	\set{y_1,z_2,y_3},\allowbreak
	\set{y_1,z_2,z_3},\allowbreak
	\set{z_1,y_2,y_3},\allowbreak
	\set{z_1,z_2,y_3},\allowbreak
	\set{z_1,z_2,z_3} }.	\\
	\end{array}$$

\item \label{53 atoms}
	An atom $A_{ij}$ 
	is satisfied by a truth value assignment $\sigma$
	iff $\rho(\sigma) \in \xi(\psi(A_{ij}))$:

	If $A_{ij} = x_j$, then
	$$\begin{array}{ll@{\hspace{1cm}}l}
	& \xi(\psi(A_{ij}))	\\
	= & \xi(d(F_1, \ldots, P_j, \ldots, F_n)) 
		& \mbox{Def.\ } \psi	\\
	= & \set{ \var(t) 
		\mid t \in \L(d(F_1, \ldots, P_j, \ldots, F_n)) }
		& \mbox{Def.\ } \xi	\\
	= & \set{ \set{ w_1, \ldots, y_j, \ldots, w_n}
		\mid \bigwedge_{k \neq j} w_k \in \set{ y_k, z_k} }
		& \mbox{Def.\ } \L, \var, F_k, P_j \\
	= & \set{ v \in V \mid y_j \in v }
		& \mbox{Def.\ } V	\\
	\multicolumn{3}{l}{\mbox{Hence,}}	\\
	& A_{ij} \mbox{ is satisfied by } \sigma	\\
	\Lra & \sigma(x_j) = {\it true} & A_{ij} = x_j	\\
	\Lra & y_j \in \rho(\sigma) & \mbox{Def.\ } \rho	\\
	\Lra & \rho(\sigma) \in \xi(\psi(A_{ij}))
		& \mbox{by the above argument}	\\
	\end{array}$$

	If $A_{ij} = (\lnot x_j)$, then
	similarly
	$\xi(\psi(A_{ij})) = \set{ v \in V \mid z_j \in v }$;
	\\
	and $\sigma$ satisfies $A_{ij}$ 
	iff $\rho(\sigma) \in \xi(\psi(A_{ij}))$.

\item \label{53 disjuncts}
	A disjunct
	$D_i$
	is satisfied by a truth value assignment $\sigma$
	iff $\rho(\sigma) \in \xi(D'_i)$:

	$$\begin{array}{@{}ll@{\hspace{1cm}}l@{}}
	& \sigma \mbox{ satisfies } D_i	\\
	\Lra & \exists j: \; \sigma \mbox{ satisfies } A_{ij}
		& \mbox{Def.\ } D_i	\\
	\Lra & \exists j: \; \rho(\sigma) \in \xi(\psi(A_{ij}))
		& \mbox{by~\ref{53 atoms}}	\\
	\Lra & \rho(\sigma) \in \bigcup_{j=1}^{n_i} \xi(\psi(A_{ij})) \\
	\Lra & \rho(\sigma) \in \xi(\; \bigmid_{j=1}^{n_i} \psi(A_{ij}))
		& \mbox{Def.\ } \xi, \L	\\
	\Lra & \rho(\sigma) \in \xi(D'_i)
		& \mbox{Def.\ } D'_i	\\
	\end{array}$$

\item \label{53 union pointwise C' D'_i}
	$\WG(C') \subseteq \xi(C') = \bigcupP_{i=1}^m \xi(D'_i)$:

	The inclusion from the definition of $\WG$.
	The equality follows
	from the definitions of $\xi$, $\L$, and $\var$.
	In the example, we have
	$\WG(C') = \set{
	\set{y_1,y_2,y_3},\allowbreak
	\set{y_1,z_2,y_3},\allowbreak
	\set{y_1,z_2,z_3},\allowbreak
	\set{z_1,z_2,z_3},\allowbreak
	\set{z_1,y_2,y_3,z_3}
	}$
	and e.g.\
	$\set{y_1,z_1,y_2,y_3} \in \xi(C') \setminus \WG(C')$.

\item \label{53 conjunct}
	$C$ is satisfied by a truth value assignment $\sigma$
	iff $\rho(\sigma) \in \xi(C')$:
	$$\begin{array}{@{}ll@{\hspace{1cm}}l@{}}
	& \sigma \mbox{ satisfies } C	\\
	\Lra & \forall i: \; \sigma \mbox{ satisfies } D_i
		& \mbox{Def.\ } C	\\
	\Lra & \forall i: \; \rho(\sigma) \in \xi(D'_i)
		& \mbox{by \ref{53 disjuncts}}	\\
	\Lra & \rho(\sigma) \in \bigcupP_{i=1}^m \xi(D'_i)
		& \mbox{by \ref{53 card disjuncts} 
		and Lem.~\eqr{55}.3}	\\
	\Lra & \rho(\sigma) \in \xi(C')
		& \mbox{by \ref{53 union pointwise C' D'_i}}	\\
	\end{array}$$

\item \label{53 least}
	Let $v$ be a member of $\WG(C')$ with least cardinality.
	$C$ is satisfiable iff $\abs{v} = n$:

	$$\begin{array}{@{}ll@{\hspace{1cm}}l@{}}
	& \abs{v} = n	\\
	\Ra & \forall i: \; v \in \xi(D'_i)
		& \mbox{by Lem.~\eqr{55}.3,
		using~\ref{53 union pointwise C' D'_i}
		and~\ref{53 card disjuncts}}	\\
	\Ra & v \in V & \mbox{by~\ref{53 card disjuncts}}	\\
	\Ra & \exists \sigma: \; v = \rho(\sigma)
		& \rho \mbox{ bijective}	\\
	\Ra & \sigma \mbox{ satisfies } C' 
		& \mbox{by~\ref{53 conjunct}}	\\ 
	\Ra & \abs{v} \leq n
		& \mbox{since } v \mbox{ has minimal cardinality in }
		\xi(C')	\\
	\Ra & \abs{v} = n
		& \mbox{by~\ref{53 union pointwise C' D'_i},
		\ref{53 card disjuncts},
		and Lem.~\eqr{55}.2}	\\
	\end{array}$$

\end{enumerate}

Hence, the NP complete
problem to decide the satisifiability of a conjunctive normal form
$C$
has been reduced to the problem to compute the minimal cardinality of 
a set from $\WG(C')$.
\qed
}

\bibliographystyle{alpha}
\bibliography{lit}


\end{document}